\documentclass[aps,prx,superscriptaddress,notitlepage,nopacs,amsmath,amstex,amssymb,citeautoscript,longbibliography,floatfix,letter]{revtex4-2}
\usepackage{graphicx} 
\usepackage{amsmath}
\usepackage{braket}
\usepackage[dvipsnames]{xcolor}
\usepackage[bookmarks=true,colorlinks,linkcolor=OrangeRed,urlcolor=NavyBlue,citecolor=RoyalBlue]{hyperref}
\usepackage{float}

\newcommand{\Pu}{\hat{P}^{\uparrow}}
\newcommand{\Pd}{\hat{P}^{\downarrow}}

\begin{document}

\title{
Stirring the false vacuum via interacting quantized bubbles \\ on a 5564-qubit quantum annealer
}

\author{Jaka Vodeb}\email{j.vodeb@fz-juelich.de}
\affiliation{Jülich Supercomputing Centre, Institute for Advanced Simulation, Forschungszentrum Jülich, 52425 Jülich, Germany}
\author{Jean-Yves Desaules}\email{jean-yves.desaules@ist.ac.at}
\affiliation{Institute of Science and Technology Austria (ISTA), Am Campus 1, 3400 Klosterneuburg, Austria}
\affiliation{School of Physics and Astronomy, University of Leeds, Leeds LS2 9JT, UK}
\author{Andrew Hallam}
\affiliation{School of Physics and Astronomy, University of Leeds, Leeds LS2 9JT, UK}
\author{Andrea Rava}
\affiliation{Jülich Supercomputing Centre, Institute for Advanced Simulation, Forschungszentrum Jülich, 52425 Jülich, Germany}
\affiliation{RWTH Aachen University, 52056 Aachen, Germany}
\author{Gregor Humar}
\affiliation{Department of Complex Matter, Jožef Stefan Institute, Jamova 39, 1000 Ljubljana, Slovenia}
\affiliation{Department of Physics, Faculty for Mathematics and Physics, Jadranska 19, University of Ljubljana, 1000 Ljubljana, Slovenia}
\author{Dennis Willsch}
\affiliation{Jülich Supercomputing Centre, Institute for Advanced Simulation, Forschungszentrum Jülich, 52425 Jülich, Germany}
\author{Fengping Jin}
\affiliation{Jülich Supercomputing Centre, Institute for Advanced Simulation, Forschungszentrum Jülich, 52425 Jülich, Germany}
\author{Madita Willsch}
\affiliation{Jülich Supercomputing Centre, Institute for Advanced Simulation, Forschungszentrum Jülich, 52425 Jülich, Germany}
\affiliation{AIDAS, 52425 Jülich, Germany}
\author{Kristel Michielsen}
\affiliation{Jülich Supercomputing Centre, Institute for Advanced Simulation, Forschungszentrum Jülich, 52425 Jülich, Germany}
\affiliation{RWTH Aachen University, 52056 Aachen, Germany}
\affiliation{AIDAS, 52425 Jülich, Germany}
\author{Zlatko Papi\'c}\email{z.papic@leeds.ac.uk}
\affiliation{School of Physics and Astronomy, University of Leeds, Leeds LS2 9JT, UK}

\date{\today}

\begin{abstract}
False vacuum decay is a potential mechanism governing the evolution of the early Universe, with profound connections to non-equilibrium quantum physics, including quenched dynamics, the Kibble-Zurek mechanism, and dynamical metastability. 
The non-perturbative character of the false vacuum decay and the scarcity of its experimental probes make the effect notoriously difficult to study, with many basic open questions, such as how the bubbles of true vacuum form, move and interact with each other. Here we utilize a quantum annealer with 5564 superconducting flux qubits to directly observe quantized bubble formation in real time -- the hallmark of false vacuum decay dynamics. Moreover, we develop an effective model that describes the initial bubble creation and subsequent interaction effects. We demonstrate that the effective model remains accurate in the presence of dissipation, showing that our annealer can access coherent scaling laws in driven many-body dynamics of 5564 qubits for over $1\mu$s, i.e., more than 1000 intrinsic qubit time units.
This work sets the stage for exploring late-time dynamics of the false vacuum at computationally intractable system sizes, dimensionality, and topology in quantum annealer platforms.
\end{abstract}

\maketitle

Nearly half a century ago, Coleman proposed the idea that our Universe may have cooled down into a metastable ``false vacuum" state after the Big Bang and the time of tunneling to the ground state or ``true vacuum" was estimated to be comparable to the lifetime of the Universe~\cite{coleman1977fate}. The idea was then further developed and applied to various cosmological observations and theories~\cite{kobsarev1974bubbles,linde1981fate,guth1981inflationary,hawking1982supercooled,abdalla2022cosmology,isidori2001metastability,degrassi2012higgs}, with ongoing attempts to observe the signatures of false vacuum decay in gravitational waves \cite{caprini2020detecting}.

The dynamics of false vacuum decay are believed to consist of ``bubbles'' of true vacuum forming in the background of false vacuum, where the size of a bubble is determined by balancing the energy gain proportional to the bubble volume and energy loss proportional to the bubble surface. Bubbles are typically assumed to undergo isolated quantum tunneling events, growing classically at a model-dependent speed \cite{caprini2020detecting}. The quantum process is difficult to study due to the non-perturbative nature of the dynamics. To circumvent this issue, early theoretical works have explored the possibility of directly creating new Universes in a laboratory setting \cite{farhi1990possible} and in engineered platforms based on condensed matter systems \cite{zurek1996cosmological}. With the advances in ultracold atomic gases, certain aspects of the false-vacuum decay dynamics can now be studied in table-top experiments \cite{zenesini2024false}.

Recently, there has been a flurry of interest in simulating quantum field theories using synthetic platforms of ultracold atoms in optical lattices, superconducting circuits, trapped ions and Rydberg atoms~\cite{Banuls2020, Bauer2023, halimeh2023coldatom}, with different proposals addressing specifically the decay of the false vacuum~\cite{billam2019simulating,billam2020simulating,abel2021quantum,ng2021fate,milsted2022collisions,Darbha2024_1,Darbha2024_2}. Two main approaches to quantum simulation involve either using quantum gates on a digital quantum computer to directly emulate the quantum field theory in question, or setting up an analogous system that exhibits a controllable first-order quantum phase transition, where it is possible to initialize in the false vacuum. In this paper, we take the latter approach and set up a quantum annealer with $5564$ superconducting flux qubits, which had previously been used to study the spin glass transition~\cite{harris2018phase} and the Kibble-Zurek mechanism \cite{bando2020probing,king2022coherent,king2023quantum}. We arrange the qubits in a ring by coupling them via ferromagnetic interactions in the presence of a transverse magnetic field, thus realizing the quantum Ising model. By then tuning the uniform longitudinal field, we initialize the system in the metastable false vacuum state and observe the decay into the true vacuum. The discrete nature of the qubit lattice gives us a direct window into the quantized bubble creation, whereby a cascade of bubble sizes is seen to emerge by tuning the longitudinal field. Moreover, the longitudinal field in the quantum annealer exhibits intrinsic modulation throughout the decay, driving the dynamics and extending the regime where we observe the same scaling laws as in coherent quantum dynamics up to 1000 qubit time units.

Quench dynamics of the Ising chain in transverse and longitudinal fields have recently attracted much interest due to the confinement effect imposed by the longitudinal field \cite{kormos2017real,liu2019confined,tan2021domain,vovrosh2021confinement,Lagnese2023}. The latter has direct implications for false vacuum decay enabling analytic predictions of the decay rate \cite{rutkevich1999decay,lagnese2021false}. Our simulation targets a different regime where quantized bubbles dominate the out-of-equilibrium dynamics, originally proposed in the context of the generalized Kibble-Zurek effect \cite{Sinha2021nonadiabiatic}. This enables us to access false vacuum decay dynamics beyond the initial bubble creation and into the previously unexplored regime of interacting bubbles. 
In contrast to the typical false vacuum decay mechanism \cite{coleman1977fate,caprini2020detecting,lagnese2021false},
we find that a large quantized bubble cannot spread in isolation. It is only through the interaction of two neighboring bubbles that one bubble can enlarge itself by reducing the size of the other. Once reduced to the smallest size of one lattice site, the bubble can then move freely along the system. These results suggest a new physical picture of the false vacuum dynamics as a heterogeneous gas of bubbles, where the smallest ``light'' bubbles bounce around in the background of larger ``heavy'' bubbles that directly interact with each other.

\section{Quantum simulation of false vacuum decay}

We study the ferromagnetic quantum Ising model in transverse and longitudinal fields on a ring with $N$ sites:
\begin{equation}
    \hat{H}=-J\sum_{j=1}^N \hat{\sigma}^z_j\hat{\sigma}^z_{j+1} - h_x \sum_{j=1}^N \hat{\sigma}^x_j - h_z\sum_{j=1}^N \hat{\sigma}^z_j,
    \label{eq:fullmodel}
\end{equation}
where $\hat{\sigma}^\alpha$ are the Pauli matrices, $J>0$ is the ferromagnetic interaction strength between nearest-neighbor spins, $h_x$ and $h_z$ are the transverse and longitudinal fields, respectively. We apply periodic boundary conditions by identifying spin $\hat \sigma_{N{+}1}^z \equiv \hat \sigma_1^z$. The field $h_x$ drives the quantum dynamics of the system, while $h_z$ imposes an energy bias between the states $\ket{\uparrow}$ and $\ket{\downarrow}$. 

\begin{figure}[ht]
    \includegraphics[width=\columnwidth]{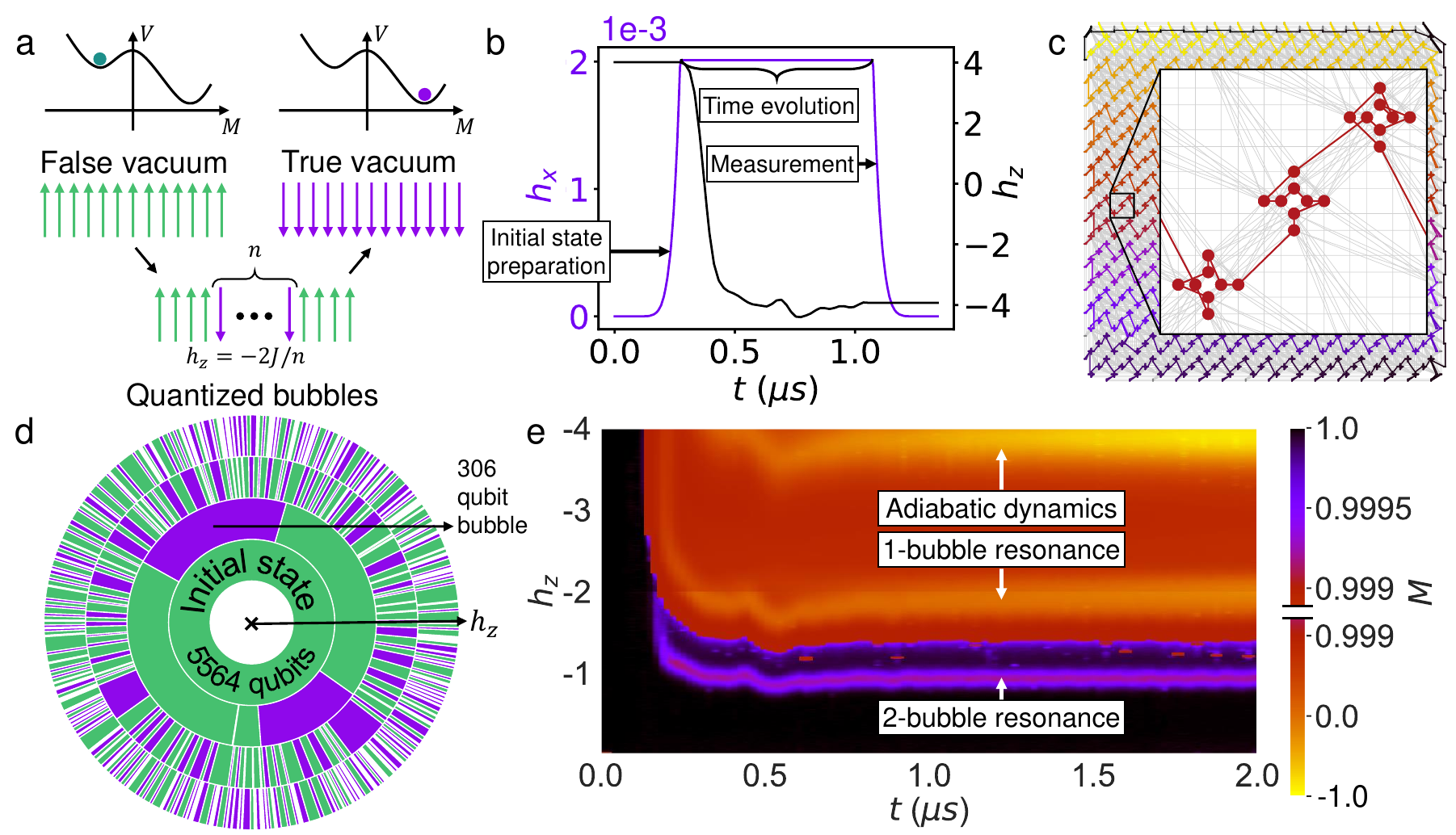}
    \centering
    \caption{
    \textbf{Realizing false vacuum decay on a quantum annealer.}
    \textbf{a} The semi-classical energy landscape $V$ as a function of magnetization $M$ of a ferromagnetic Ising chain in a transverse $h_x$ and longitudinal $h_z$ field. The landscape exhibits a local metastable minimum dubbed as the false vacuum, represented by the polarized $\ket{\uparrow\uparrow\cdots \uparrow}$ state. The global minimum or true vacuum is the other polarized $\ket{\downarrow\downarrow\cdots\downarrow}$ state. The false vacuum decay unfolds via creation of quantized true-vacuum bubbles of size $n$, determined by the energy balance between the surface ($4J$) and volume energy contributions ($2h_zn$).
    \textbf{b} False vacuum decay observation protocol. We initialize all qubits in the $\ket{\uparrow\uparrow\cdots\uparrow}$ state by setting $h_z>0$ and adiabatically switch $h_x$ from $0$ to a small value $h_x\ll J$ over $10\mu s$ ($0.27\mu s$ is used in the plot for clarity purposes). Then we flip the sign of $h_z$, swapping the true and false vacuum states, and observing the dynamics for a time duration $t$. Finally, we turn $h_x$ back to $0$ as fast as possible ($\geq0.18$ $\mu s$) and measure the spin configuration in the $\hat \sigma^z$ basis. This protocol is repeated $1000$ times for each value of $t$.
    \textbf{c} Embedding of a $5564$-qubit ring on the Pegasus graph of the $5614$-qubit device D-Wave $Advantage\_system5.4$. The Pegasus graph contains $15\times15\times3$ $8$-qubit Chimera cells with complete bipartite connectivity (colored crosses) that are coupled by additional external and odd couplers (gray lines)~\cite{dwave2020TechnicalDescription}, such that each qubit is connected to 15 other qubits on average. 
    Qubits within the $8$-qubit cells are connected along randomly sampled one-dimensional chains (inset).
    \textbf{d} Spin configurations measured in our quantum simulation. The inner ring shows the initial false vacuum state comprised of $5564$ spins (for clarity, only $1000$ out of $5564$ spins in a single configuration are shown). The outer 3 rings show configurations measured at $h_z=-0.1,-0.5,-2$ with $h_z$ decreasing radially. An example of a large $n=306$ quantized bubble shown in purple highlights the extent of the observed bubble sizes.
    \textbf{e} Magnetization $M$ heat profile versus time $t$ and longitudinal field magnitude $h_z$ at transverse field strength $h_x=0.002$. 
    The color scheme is split into two separate linear scales, a larger scale from $-1$ to $0.999$ (bottom half) and a smaller scale from $0.999$ to $1$ (upper half). The adiabatic dynamics and the $n=1$-bubble resonance are easily observed on the larger scale, while the $n=2$-bubble resonance can only be resolved in the 4th decimal of $M$, due to the decrease of the rate of dynamics by an order of magnitude. The apparent resonance at $h_z=-4$ is identified with adiabatic dynamics rather than bubble creation, where the system follows an instantaneous ground state during the evolution.
    }
    \label{fig:Fig1}
\end{figure}

In the regime $0 \leq h_x\ll J$ and $h_z=0$, there are two degenerate ground states $\ket{\uparrow ...\uparrow}$ and $\ket{\downarrow ...\downarrow}$. When $h_z>0$, the $\ket{\uparrow ...\uparrow}$ state becomes the ground or true vacuum state and $\ket{\downarrow ...\downarrow}$ a metastable or false vacuum state, see Fig.~\ref{fig:Fig1}a. By first setting $h_z>0$ and adiabatically turning on $h_x$ to a small value $h_x\ll J$, we initialize the system in the $\ket{\uparrow ...\uparrow}$ product state. Then we induce a first-order quantum phase transition by flipping the sign of $h_z$, swapping the true and false vacuum states, and observe the dynamics for a time duration $t$. Finally, we turn $h_x$ back to zero as fast as possible and measure the spin configuration in the $\hat \sigma^z$ basis. Fig. \ref{fig:Fig1}b illustrates the described protocol, while Fig. \ref{fig:Fig1}c shows the embedding of the spin chain in a qubit array used in our quantum simulations. We note here that $h_z(t)$ was determined experimentally through single-qubit measurements and exhibits large modulation around the final target value after the flip. This modulation extends up to $t\sim0.75\mu s$ in the evolution time and it will play an important role in the interpretation of our data.

Our quantum simulations are performed in the small $h_x\ll J$ regime, where we can apply  semiclassical intuition based on the diagonal part of the Hamiltonian in the $z$-basis. In this case, it is useful to gather possible configurations of the system into sectors with the same value of magnetization, $M=\langle\sum_i\hat \sigma_i^z/N\rangle$, separated by energy gaps determined by the value of $h_z$. For general values of $h_z$, the initial $\ket{\uparrow ...\uparrow}$ state stays an eigenstate in its own $M$-sector after the $h_z$ sign flip and no dynamics of $M$ are observed. This is due to the large energy separation between different $M$ sectors that cannot be hybridized by a small $h_x$. However, for specific values of $h_z=-2J/n$, where $n>0$ is an integer, the surface energy cost for flipping a domain of $n$ spins, $4J$, is exactly balanced out by the volume energy gain, $2h_zn$~\cite{Sinha2021nonadiabiatic}. Hence, an arbitrarily small $h_x$ is sufficient to hybridize the classical computational basis states into eigenstates consisting of a superposition of the $\ket{\uparrow...\uparrow}$ state and so-called $n$-bubbles, i.e., domain walls in the background of $\ket{\uparrow ...\uparrow}$. For example, $\ket{\uparrow\uparrow\uparrow\downarrow\downarrow\downarrow\uparrow\uparrow\uparrow}$ is a state with a single 3-bubble. Fig.~\ref{fig:Fig1}d shows the spin configurations measured in our quantum simulations with bubble sizes up to $306$ spins, which is consistent with the theoretical prediction in which we can form increasingly larger bubbles by decreasing $h_z$ according to $h_z=-2J/n$. For these discrete values of $h_z$, the initial state is no longer an eigenstate and undergoes nontrivial quantum dynamics, resulting in large changes in $M$.  Fig.~\ref{fig:Fig1}e shows the observed $n=1$ and $n=2$ resonances, where large changes in $M$ can be seen at $h_z=-2J$ and $h_z=-J$, respectively, in contrast to other values of $h_z$.

We note that significant changes in $M$ can also be observed in Fig.~\ref{fig:Fig1}e at values of $h_z\approx -4J$, where no dynamical resonances are expected. Such a large magnitude of $h_z$ leads to thermally assisted adiabatic dynamics~\cite{amin2008thermally}, where the system can follow the instantaneous ground state during time evolution. The adiabatic theorem is applicable if the time scale of Hamiltonian changes is slower than or comparable to $t_a\propto\Delta_{min}^{-2}$, where $\Delta_{min}$ is the minimum gap between the instantaneous ground and first excited state. In the case $|h_z|\gg J, h_x$, the gap becomes large enough for the time scale of $h_z(t)$ to match $t_a$. Therefore, no bubble creation takes place and the spins turn simultaneously and in accordance with $h_z(t)$ initially, changing the state from fully polarized and triggering more complex resonant processes, see Supplementary Section 2.

\section{Observation of quantized bubbles and dynamical scaling laws}

\begin{figure}[htb]
    \includegraphics[width=\columnwidth]{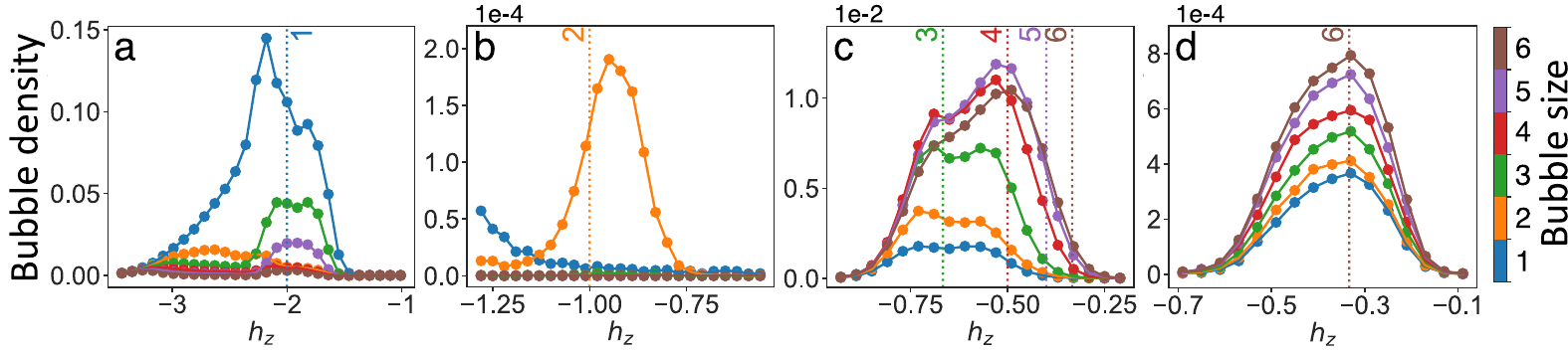}
    \centering
    \caption{
    \textbf{Observation of quantized bubbles.}
    \textbf{a}-\textbf{d} Bubble density measurements at $J=1$ and different $h_z$ magnitudes, with $h_x=0.002$, $t=2\mu s$ in \textbf{a,b}, $h_x=0.05$, $t=1\mu s$  in \textbf{c}, and $h_x=0.1$, $t=1\mu s$ in \textbf{d}.  
    The bubble sizes $n=1,2,\ldots, 6$ are seen to be dominant around their respective resonances $h_z=-2J/n$, indicated by vertical dotted lines. Error bars across the entire figure are smaller than the size of the symbols. 
    }
    \label{fig:Fig2}
\end{figure}

To ascertain which bubbles are involved in magnetization changes, we measured the $n$-bubble density 
$\lambda_n=(1/N)\sum_{i=1}^N\braket{\Pu_i[\prod_{j=1}^n\Pd_{i+j}]\Pu_{i+n+1}}$, where $\hat P^\sigma = \ket{\sigma}\bra{\sigma}$ is a projector on the local $\sigma=\uparrow,\downarrow$ spin state. Figures \ref{fig:Fig2}a-d show the detected $n=1,2,3,4,5,6$-bubble resonances. We observe a strong suppression of all other bubble sizes except the expected ones. According to the theoretical analysis presented in Methods, the leading-order effective Hamiltonian describing an $n$-bubble resonance is proportional to $h_x^n$. If we assume $h_x<J$, 1-bubbles are the fastest, then 2-bubbles, etc., arbitrarily slowing down the dynamics as $n$ increases. Figures~\ref{fig:Fig2}a-d show that we need to increase $h_x$ by at least two orders of magnitude to begin to observe hints of higher resonances through low-density bubble formation, which is consistent with the theoretical prediction. 

\begin{figure}[bt]
    \includegraphics[width=\columnwidth]{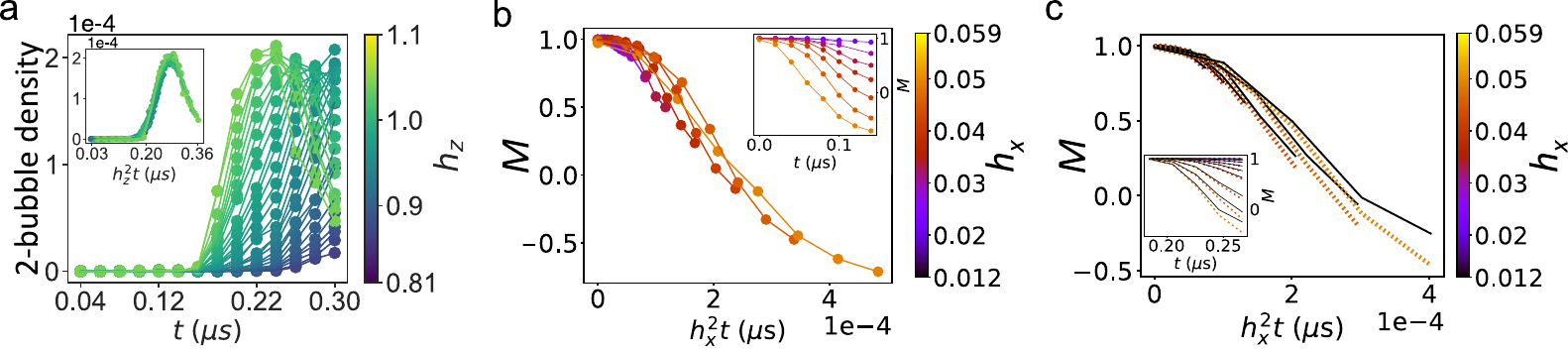}
    \centering
    \caption{
    \textbf{Scaling laws for bubble dynamics.}
    \textbf{a} 2-bubble density at $h_x=0.002$ as a function of time at various $h_z$ magnitudes (color bar). Inset shows the collapse of different curves when time is rescaled by $h_z^2$ in accordance with Landau-Zener theory~\cite{Sinha2021nonadiabiatic}.
   \textbf{b}-\textbf{c} Magnetization at $h_z=-J$ resonance as a function of rescaled time $h_x^2 t$, for different values of $h_x$ indicated on the color bar. Both the measured magnetization curves in {\bf b} and the 3-spin Bloch-Redfield numerical simulations in {\bf c} follow the same $h_x^2$ scaling law, suggesting that the effective Hamiltonian governing the dynamics is proportional to $h_x^2$. The inset in {\bf b} shows the raw data obtained on the quantum annealer without rescaling.  The unscaled results of the numerical simulations are shown in the inset of {\bf c}, where the black curves show the magnetization in the effective model describing the $h_z=-J$ resonance (see Methods). Error bars across the entire figure are smaller than the size of the symbols.
    }
    \label{fig:Fig3}
\end{figure}

In a two-level approximation~\cite{Sinha2021nonadiabiatic}, tunneling events to different $n$-bubbles can be thought of as Landau-Zener transitions, where the metastable state $\ket{\uparrow ...\uparrow}$ and an $n$-bubble state at the appropriate resonant conditions are the two states involved in the anticrossing. According to Landau-Zener theory, it follows that $n$-bubble density $\lambda_n\propto\tau_Q h_x^n$ should be proportional to the product of the time it takes for the Hamiltonian to traverse the anticrossing $\tau_Q$, determined by $h_z(t)$ in our case, and the $n$-th power of $h_x$. Using our single-qubit measurements we show that the time it takes for $h_z(t)$ to reach zero during its sign flip is proportional to the square of its magnitude ($\tau_Q\propto \vert h_z\vert^2$), see Supplementary Section 3 for details. This means that $\lambda_n(t)$ curves measured at different pause times between the initialization and measurement ramp $t$ should collapse onto a single curve if we multiply $t$ by $h_z^2$.  Figure~\ref{fig:Fig3}a shows that the $\lambda_2$ curves indeed exhibit a collapse according to this law.

Nevertheless, in order to fully understand the dynamics in our quantum annealer, it is necessary to account for all dominant processes and not only the creation of bubbles. We will focus on the dynamics from the initial state $\ket{\uparrow \cdots \uparrow}$, for which the creation of $n$-bubbles happens at $h_z=-2J/n$. For each resonance, we have derived the corresponding effective model using the Schrieffer-Wolff transformation~\cite{Bravyi2011} and we present the effective Hamiltonians at leading orders in the Methods section. 
The effective Hamiltonian describing the dynamics at the 2-bubble resonance at $h_z=-J$ is proportional to $h_x^2$. Figure~\ref{fig:Fig3}b shows magnetization measurements taken at this resonance using the quantum annealer and how the $M(t)$ curves collapse when scaling the time axis with $h_x^2$. Our numerical emulation of the quantum annealer in Fig.~\ref{fig:Fig3}c suggests that the $h_x^2$ scaling law is the same as in the case of coherent quantum dynamics. Figure~\ref{fig:Fig3}b shows only the initial behavior of $M(t)$, which follows the $h_z(t)$ modulation at later times; however, after the $h_z(t)$ modulation stops, an $h_x^3$ scaling law emerges as a consequence of thermalization combined with a relatively slow quantum simulation measurement ramp, see Supplementary Sections 6-9 for more details.

\section{Bubble interactions}

\begin{figure}[b]
    \includegraphics[width=\columnwidth]{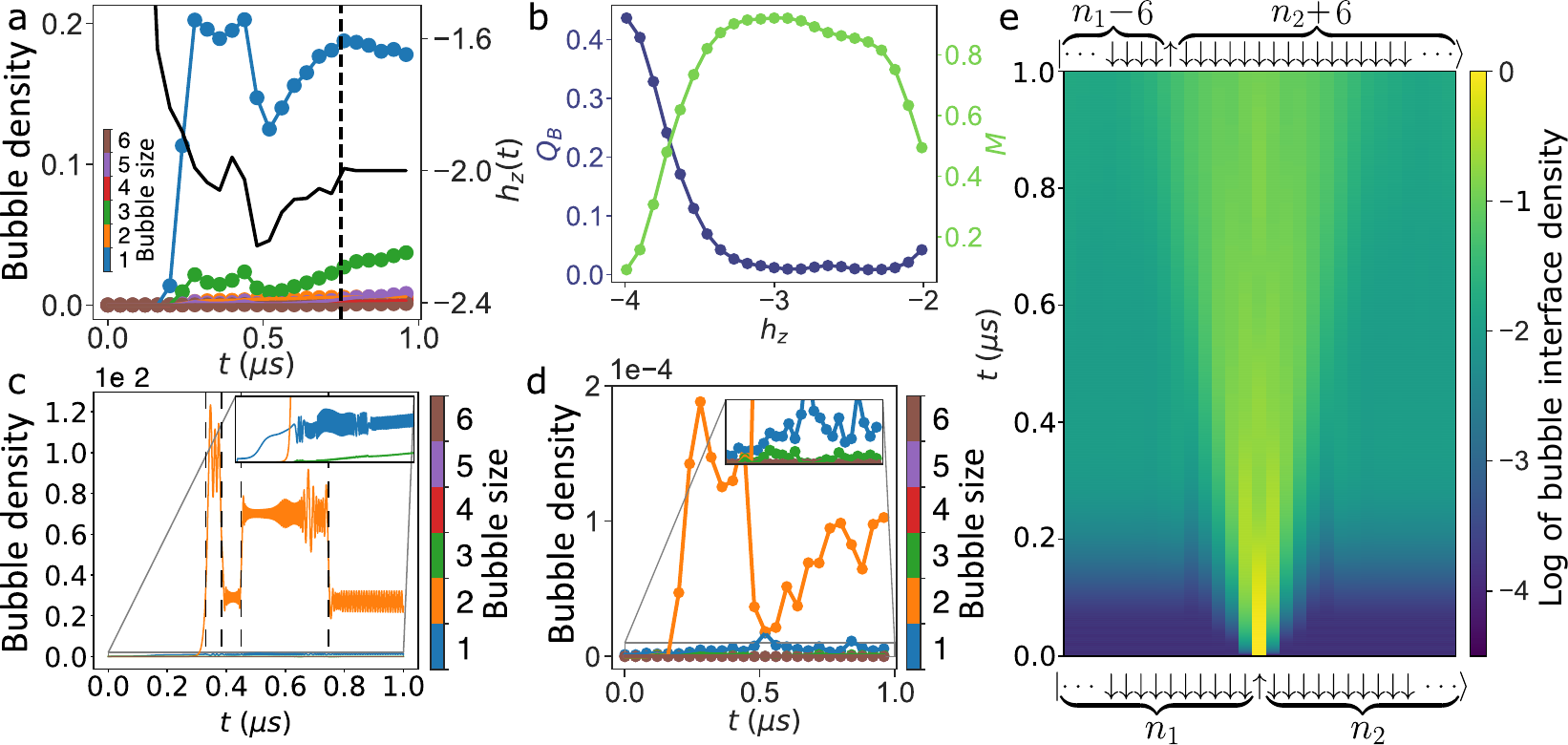}
    \centering
    \caption{
    \textbf{a}
    Time series measurements of the bubble density at the 1-bubble resonance $h_z=-2J$ and $h_x=0.002$. 
     During the initial $h_z(t)$ modulation, the profile of which is shown by the black curve on the right axis, the 1-bubble density (color bar) governs the dynamics.
      After about $\sim 0.75\mu s$ (dashed line),  thermalization effects take over by transforming 1-bubbles into 3- and 5-bubbles. 
    \textbf{b} Measurement of the emergent blockade $Q_B$ and magnetization $M$ (right axis) at $h_x=0.002$ and $t=0.38\mu s$, plotted as a function of $h_z$ magnitude. The blockade condition is violated (deviates from 0) only at $h_z$ values significantly off the 1-bubble resonance ($h_z\leq-3.5$), accompanied by large changes in $M$. Near the resonance ($h_z\approx-2$), even though large changes in $M$ occur, the blockade condition is respected.
    \textbf{c}-\textbf{d} 
    Dynamics at the resonance $h_z=-J$ with fixed $h_x=0.0203$ and $h_x=0.002$, respectively.
    Matrix-product state simulation with 100 qubits in \textbf{c} captures some of the key aspects of the data obtained on the quantum annealer in {\bf d}. The black dashed lines in \textbf{c} show the sudden change in the number of 2-bubbles when $h_z(t)$ is exactly at the resonance point. The inset magnifies the low-density regime, where only 1- and 3-bubbles can be seen. The increase of 1- and 3-bubbles at later times is likely due to 2-bubbles interacting. The quantum simulation using $h_x=0.002$ in \textbf{d} shows good agreement with the theoretical prediction in {\bf c}. For all the measured data, error bars are smaller than the symbol size.
    \textbf{e} Matrix-product-state simulation of the dynamics after an instantaneous quench from a product state shown at the bottom, containing two large bubbles ($n_1=23$ and $n_2=24$ spins) next to each other in a system with 50 spins in total. The system undergoes coherent evolution with fixed $h_z=-1$ and $h_x=0.02$, and the color bar shows the ``bubble interface density'', $\langle \hat P_{j-1}^\downarrow \hat P_j^\uparrow \hat P_{j+1}^\downarrow \rangle$, on a log scale and for all sites $j$. The moving front corresponds to the two bubbles exchanging $\downarrow$-spins and changing their sizes. The final state at the end of the evolution is a quantum superposition, with one of the classical configurations shown at the top. 
    }
    \label{fig:Fig4}
\end{figure}

The measured dynamics of different bubble sizes at $h_z=-2J$ resonance in Fig.~\ref{fig:Fig4}a is consistent with the picture that 1-bubbles remain approximately quantized and do not grow with time. On the quantum annealer, this persists until thermalization kicks in and 1-bubbles start to transform into 3- and 5-bubbles, with 2- and 4-bubbles remaining suppressed throughout the time evolution. The exploration of this peculiar thermalization effect is beyond the scope of this work, as thermalization and bubble interaction effects cannot be easily separated from each other in our quantum annealer due to decoherence effects. Nevertheless, we now argue that bubble interactions play a crucial role in the dynamics at higher $n>1$ resonances if a system is perfectly isolated from the environment. We will demonstrate this in the framework of effective models, presented in Methods and previously mentioned in the context of Fig.~\ref{fig:Fig3}. 

At the $n=1$ resonance, 1-bubbles can be created with a rate proportional to $h_x$. These 1-bubbles can then hop along the chain with a rate proportional to $h_x^2/J$, but they cannot merge with each other to create large bubbles. In fact, even when accounting for higher-order processes, there is still no path to create $(n>1)$-bubbles when starting from the $\ket{\uparrow\uparrow\uparrow\cdots}$ state. This dictates that, at the $n=1$ resonance, there can never be two $\downarrow$-spins next to each other. The system therefore experiences an emergent kinetic constraint, reminiscent of the Rydberg blockade phenomenon~\cite{Bernien2017Rydberg}. We quantify the blockade by measuring the operator
  $\hat{Q}_{B}=(1/N)\sum_j \Pd_{j}\Pd_{j+1}$,
which counts the density of neighboring $\downarrow$-spins. We expect $\langle \hat{Q}_{B}\rangle $ to be strongly suppressed around $h_z=-2J$, rising towards 0.5 in other dynamical settings. Figure~\ref{fig:Fig4}b shows a good match between these predictions and quantum simulation data. Meanwhile, total magnetization strongly deviates from the initial value of $1$, showing that the lack of neighboring excitations is not trivially due to frozen dynamics.

Now we consider higher-order resonances $h_z=-2J/n$, with $n>1$, where bubbles contain $n$ spins and are created at a rate proportional to $h_x^n/J^{n-1}$. Once these large bubbles are created, they cannot hop around. However, bubbles can now take or give $\downarrow$-spins to neighboring bubbles, allowing them to change size. This occurs with a rate  $\propto h_x^2/J$ which, for large $n$, is much faster than bubble creation. This can also lead to a bubble shrinking all the way down to a 1-bubble, which can then hop along the chain. This means that, despite large bubbles being stuck in place, information can still flow through inter-bubble interactions and movement of 1-bubbles.

Our theoretical predictions imply that, in a quantum simulation tuned to a $n>1$ resonance, the size of bubbles is not limited to $n$, even if the system is perfectly isolated from the environment. 
This can be seen in a fully coherent matrix-product state simulation of a system with $N=100$ spins at $n=2$ resonance in Fig.~\ref{fig:Fig4}c. While 2-bubbles dominate, 1- and 3-bubbles are also visible. This is expected as they are produced by interactions of 2-bubbles. Qualitatively similar behavior is also seen in the quantum simulation data in Fig.~\ref{fig:Fig4}d.

The data in Figs.~\ref{fig:Fig4}c-d also highlights another important property for $n>1$: the number of 2-bubbles changes abruptly at some times, while staying approximately constant during the rest of the simulation. The timings of abrupt changes coincide exactly with $h_z(t)$ hitting the appropriate resonant value, while the rest of the time the system is slightly away from the resonance. This highlights the sensitivity to the detuning, $\delta=h_z+2J/n$, which competes with $h_x^n/J^{n-1}$. As $h_x/J\ll 1$, even a small $\delta$ is enough to overpower the bubble creation terms for $n>1$. As the detuning is a diagonal contribution, it leads to the suppression of all dynamical processes, including bubble creation. This pattern of sudden changes due to the fluctuation of $h_z$ is clearly captured in the numerical simulation in Fig.~\ref{fig:Fig4}c but it is also visible in the annealer data in Fig.~\ref{fig:Fig4}d. We note that this sensitivity to detuning is expected to be less strong for $n=1$, as in that case $\delta$ only competes with $h_x$.

To further highlight the importance of bubble interactions, we have studied a closed system with two large bubbles next to each other and essentially occupying the entire system, see Fig.~\ref{fig:Fig4}e. This setup leaves no room for new bubbles to appear, and the only resonant process left is the exchange between the two bubbles. We can then track the interface between them by measuring $\hat{P}_{j-1}^\downarrow \hat{P}_{j}^\uparrow\hat{P}_{j+1}^\downarrow$, which is plotted on a log scale in Fig.~\ref{fig:Fig4}e at the $h_z=-J$ resonance. While the interface density is 1 at a single location at time $t=0$ and zero everywhere else, as time goes on the interface steadily delocalizes due to the bubbles exchanging $\downarrow$-spins and thereby changing their sizes. We expect similar behavior to hold at other $n>1$ resonances.

\section{Discussion and outlook}

We have performed a quantum simulation of the false vacuum decay and identified its underlying mechanism -- the formation of quantized bubbles of true vacuum.  The large size of our 5564-qubit quantum annealer allowed us to observe considerable bubble sizes comprising up to 300 spin flips, confirming the standard cosmological scenario where the size of the formed bubble is determined by the competition between the volume energy gain and surface energy loss. Our central finding is that interactions between bubbles are the key next-order effect after bubble creation, and we have argued that their understanding is crucial for a comprehensive description of of false vacuum decay. Previous studies of false vacuum decay in quantum spin chains~\cite{rutkevich1999decay,lagnese2021false} have explored a different parameter regime where $h_x$ is not small and the energy spectrum forms a continuum. While the possibility of resonances was pointed out as a subleading effect~\cite{rutkevich1999decay}, these analytical considerations still assume a dilute bubble picture, neglecting interactions between bubbles. 
Our results therefore call for a deeper understanding of the interaction effects between bubbles, not only in microscopic models such as the one studied here, but also in quantum field theory approaches, including cosmological models of the Big Bang. 

More broadly, our work showcases that current quantum annealing devices can be useful in probing complex many-body dynamics, e.g., as demonstrated here through external field modulation on a scale of $1000$ individual qubit time units. Extensions of our model to two or three spatial dimensions with various lattice topologies are, in principle, straightforward on the same type of device, potentially reaching intractable computational complexity with a multitude of implications. Let us mention a few examples of other interesting non-equilibrium phenomena that can be accessed in the platform established here. False vacuum decay, as a specific instance of a first-order quantum phase transition, allows to probe generalizations of the Kibble-Zurek scaling laws~\cite{polkovnikov2011colloquium,kibble1976topology,zurek1996cosmological,dziarmaga2005dynamics} in such transitions, in particular comparing the predictions for the rate of defect formation after crossing the transition with quantum simulations.  Quantum metastability -- the cornerstone of the false vacuum decay phenomenon -- also underlies reaction rate theory~\cite{langer1969statistical,affleck1981quantum,caldeira1981influence,leggett1984quantum,leggett1987dynamics,hanggi1990reaction}, allowing the use of quantum simulation for estimating the transition rate of decay processes from a metastable minimum to a lower energy state in the presence of temperature, which is challenging to compute otherwise.  In the regime of stronger longitudinal fields, confinement effects are expected to become important, possibly localizing bubbles in space and giving rise to an emergent prethermalization regime~\cite{birnkammer2022prethermalization}.  
Finally, at the 1-bubble resonance, our model displays an emergent kinetic constraint that maps exactly to the so-called PXP model~\cite{FendleySachdev,Lesanovsky2012} (see Methods), which hosts quantum many-body scars~\cite{Bernien2017Rydberg, TurnerNature, Su2023TBH}, and possibly other types of ergodicity breaking, such as Hilbert space fragmentation and many-body localization, in higher dimensions~\cite{balducci2022localization,Hart2022}. 
This opens the way to probing non-ergodic phenomena in large systems in the presence of dissipation and potentially new types of scars in constrained models at other $n>1$ resonances.

\section{Methods}

\subsection{Quantum simulation on D-Wave's quantum annealer}

Our quantum simulations utilized D-Wave's quantum annealing device $Advantage\_system5.4$, which features $N_q= 5614$ qubits and is kept at a cryostat temperature of $16.4\pm 0.1 \mathrm{mK}$. The annealer implements the Hamiltonian
\begin{equation}
    \begin{aligned}
        \hat{H}_\mathrm{DW} &= -\frac{A(s)}{2}\left(\sum_{i=1}^{N_q}\hat \sigma_i^x\right) +  \frac{B(s)}{2}\left(g(t)\sum_{i=1}^{N_q}h_i\hat \sigma_i^z + \sum_{i<j}^{N_q} J_{ij}\hat \sigma_i^z\hat \sigma_j^z\right),
    \end{aligned}
\end{equation}
where $\hat \sigma_i^{x,z}$ are the Pauli matrices for the $i$th qubit, $h_i$ is the longitudinal external field at qubit $i$, $J_{ij}$ are the couplings between qubits $i$ and $j$, which are non-zero and user-tunable only if they are physically connected in the quantum processing unit (Fig. \ref{fig:Fig1}c). $A(s)$ and $B(s)$ represent the energy scales of their respective terms and are driven in time by the annealing schedule $s(t)$, which is linearly interpolated from a series of user-specified points $[(t_i,s_i)]$.

Finding a ring embedding in a given graph is an instance of an NP-complete Hamiltonian circuit problem~\cite{Karp1972KarpsNPCompleteProblems}. We generate our ring embedding on $5564$ qubits of the $Advantage\_system5.4$ graph by first connecting all $8$-qubit Chimera cells in the Pegasus topology, see Fig. \ref{fig:Fig1}c. We start in the upper-left corner and proceed horizontally, changing the horizontal direction at the end of every row, until we reach the bottom-right corner. The chain of qubits within each $8$-qubit Chimera cell is chosen along a random suitable path, see Fig. \ref{fig:Fig1}c inset. The ring is closed by proceeding along the outer qubits at the right and top edge of the graph (black part of the chain in Fig. \ref{fig:Fig1}c). This procedure yields a ring of $5446$ qubits. We then iteratively add qubits to the chain from the set of omitted remaining qubits by adding detours into the ring until we obtain the final $5564$-qubit closed chain. We note that a few of the qubits and couplers in the full Pegasus graph are not present on the device due to fabrication defects; these are accounted for individually.

We are interested in probing the dynamics of $\hat{H}$ in Eq.~(\ref{eq:fullmodel}) at a certain value of $h_x$.  Therefore, we specify the annealing schedule as $[(0,1), (irt,s_{h_x}), (irt+pt,s_{h_x}), (irt+pt+mt,1)]$, where $s_{h_x}$ is obtained from the relation $h_x=A(s_{h_x})/B(s_{h_x})$, obtained by rewriting $\hat{H}_{DW}$ as $(B(s)/2)\hat{H}$. We choose uniform $h_i=h$, $J_{ij}=-1$ and define $h_x= A(s)/B(s)$, $h_z=-g(t)h$. 
At time $t=0$, we specify the initial state for all qubits as the product state $\ket{\uparrow ... \uparrow}$. Then, within the initial ramp time $irt$, we bring the system to the desired $h_x$ value, which drives the dynamics we are interested in, and keep it constant for time $pt$. We replace $pt$ with $t$ in all plots. Finally, we bring $h_x$ to $0$ within time $mt$, which constitutes a measurement. Only after $h_x$ is brought back to $0$ is it possible to read out the state of the qubits in the computational or $\hat \sigma^z$ basis. 

Typical time scales that we used on the D-Wave device are $irt=10\mu s$, $mt=272ns$, and $pt$ ranging from $0$ to $2\mu s$. After the initial state preparation, the system remains in the $\ket{\uparrow ... \uparrow}$ state due to the small values of $h_x$ compared to $h_z$. During the entire time evolution, which lasts for a time $irt+pt+mt$, the system is subject to open system dynamics, governed by two main effects; measurement by the environment and thermalization. Our single spin measurements show that measurement by the environment is dominant whenever the system is being driven by the longitudinal external field $h_z$. Whenever $h_z$ becomes constant, thermalization effects become more evident and are heavily dependent on the value of $h_x$, which drives the quantum dynamics of the system -- see Supplementary Sections 7 and 9.

\subsection{Simulations of thermalization dynamics}

To capture thermalization effects on the system's dynamics, we employed the Bloch-Redfield master equation \cite{breuer2002theory}
\begin{equation}
    \frac{d}{dt}\rho_{ab}(t)=-i\omega_{ab}\rho_{ab}(t)+\sum_{cd}^{sec}R_{abcd}\rho_{cd},
\end{equation}
where $\hat \rho$ is the density matrix of the system, $\omega_{ab}\equiv \omega_a - \omega_b$ with $\omega_a=E_a/\hbar$ and $E_a$ being the eigenenergies of the system. $sec$ denotes the secular approximation, which states that we can neglect all fast-rotating terms in the sum, and $R_{abcd}$ is the Bloch-Redfield tensor~\cite{breuer2002theory}
\begin{equation}
    \begin{aligned}
        R_{abcd} = & -\frac{1}{2\hbar^2}\sum_{\alpha,\beta} \Biggl\{ \delta_{bd}\sum_n A_{an}^{\alpha}A_{nc}^{\beta}S_{\alpha\beta}(\omega_{cn})- A_{ac}^{\alpha}A_{db}^{\beta}S_{\alpha\beta}(\omega_{ca}) + \delta_{ac}\sum_n A_{dn}^{\alpha}A_{nb}^{\beta}S_{\alpha\beta}(\omega_{dn}) - A_{ac}^{\alpha}A_{db}^{\beta}S_{\alpha\beta}(\omega_{db}) \Biggr\},
    \end{aligned}
\end{equation}
where $A_{ab}^{\alpha}$ are the matrix elements in the system's eigenbasis of the operator $A^{\alpha}$ that couples bilinearly to the bath. Here we choose $A^{\alpha}=\sigma_{\alpha}^z$, where $\alpha$ runs through all the spins of the system. $S_{\alpha\beta}(\omega)=\eta\omega\theta(\omega)\exp{(\omega/\omega_c)}$ is the noise power spectrum of the bath, chosen to be Ohmic in our case, where $\theta(\omega)$ is the Heaviside step function, $\eta$ the coupling strength of the system-bath coupling that ranges from $0.1$ to $0.2$ in our case, and $\omega_c$ a cutoff frequency higher than any other relevant energy scale. 

The numerical simulations in Figs.~\ref{fig:Fig4}c,e were performed under the assumption of a closed system using matrix-product state (MPS) formalism \cite{Schollwock}. For efficiency, the simulated system has open boundary conditions, but we discard the boundary sites when computing observable expectation values in order to minimize boundary effects. To reach the long times required for the simulation, a 4th-order time-evolving block decimation (TEBD4) was used \cite{Vidal2003,Paeckel_2019}. For Fig.~\ref{fig:Fig4}c, the timestep is $\delta t=0.01$ while the maximum MPS bond dimension is $\chi=128$, which was never saturated during the simulation. For Fig.~\ref{fig:Fig4}e, the timestep is t=0.05 while the maximum bond dimension is $\chi=200$.

\subsection{Effective models at different resonances}

To fully understand the dynamics beyond bubble creation in the vicinity of resonances, we have derived the corresponding effective Hamiltonians using the Schrieffer-Wolff transformation~\cite{Bravyi2011}. We quote the main results here, while the derivation and detailed analysis of the models are provided in Supplementary Section 5. For $n=1$, in the sector containing the state $\ket{\uparrow\cdots \uparrow}$, the combined effective Hamiltonian at first and second order reads:
\begin{equation}\label{eq:Heff_1}
    \hat{H}^{(1,2)}_{\mathrm{eff}, n{=}1} = -h_x\sum_{j=1}^N \Pu_{j-1}\hat{\sigma}^x_j\Pu_{j+1}-\delta\sum_{j=1}^N\hat{\sigma}^z_j
    +\frac{h_x^2}{4J}\Big[\sum_{j=1}^{N}\Pu_{j-1}(\hat{\sigma}^+_j\hat{\sigma}^-_{j+1}+\hat{\sigma}^-_j\hat{\sigma}^+_{j+1})\Pu_{j+2}+2\sum_{j=1}^N \Pd_j-\frac{3}{2}\sum_{j=1}^{N}\Pd_{j-1}\Pd_{j+1} \Big],
\end{equation}
where $\delta=h_z+2J$ is the (weak) detuning away from the $n=1$ resonance, $\hat \sigma^\pm = (\hat\sigma^x \pm i \hat\sigma^y)/2$ are the standard spin raising and lowering operators, and  $\Pd=\ket{\downarrow}\bra{\downarrow}$,  $\Pu=\ket{\uparrow}\bra{\uparrow}$ are local spin projectors.

\begin{figure}
    \centering
    \includegraphics[width=0.5\linewidth]{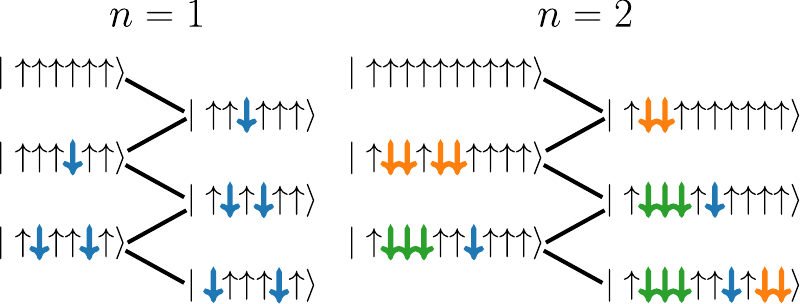}
    \caption{Schematic of the allowed dynamical processes at $n=1$ and $n=2$ resonances. At $n=1$, the 1-bubbles can be directly created and move along the chain. However, they cannot merge and their size is fixed. By contrast, for $n>1$ (here illustrated for $n=2$) only $n$-bubbles can form. While these larger bubbles cannot move, neighboring bubbles can exchange $\downarrow$-spins to change their size. If 1-bubbles are formed that way, they can hop along the chain unlike bigger bubbles. }
    \label{fig:eff}
\end{figure}

The dynamics generated by the Hamiltonian in Eq.~\eqref{eq:Heff_1} can be understood as follows. The $\Pu\hat{\sigma}^x\Pu$ term allows the creation of single-site bubbles (i.e., single $\downarrow$-spins in a background of $\uparrow$-spins), while the $\Pu\left(\hat{\sigma}^+\hat{\sigma}^-+\hat{\sigma}^-\hat{\sigma}^+\right)\Pu$ allows these bubbles to hop around. A sequence of allowed processes is illustrated in Fig.~\ref{fig:eff}. Importantly, due to the projectors, the bubbles cannot merge to form larger ones. This is also impossible to do using higher-order processes.
A simple argument is that there are no states with larger bubbles at the same classical energy (i.e., the energy contribution of the $\hat \sigma^z$ terms) as the $\ket{\uparrow\uparrow\uparrow\cdots}$ state, so it is impossible to reach these states resonantly. 

In the main text, we have demonstrated that one measurable consequence of the effective Hamiltonian in Eq.~\eqref{eq:Heff_1}  is a robust emergent kinetic constraint reminiscent of the Rydberg blockade~\cite{Bernien2017Rydberg}. 
The quality of this emergent blockade can be assessed using the operator $\hat{Q}_{B}$ introduced in the main text, which measures the density of neighboring $\downarrow$-spins and can be equivalently expressed in the spin language as
$\hat{Q}_{B} = 1/4 + (1/(4N))\sum_j \hat{\sigma}^z_j\hat{\sigma}^z_{j+1}-(1/(2N))\sum_j\hat{\sigma}^z_j$. 

For $n>1$ resonances, the bubble creation term is no longer dominant as it happens at order $n$ according to
\begin{equation}
    \hat{H}_{\mathrm{eff},n} = c_n\frac{h_x^n}{J^{n-1}}\sum_{j=1}^N\Pu_j\left(\prod_{k=1}^n\hat{\sigma}^-_{j+k}\right)\Pu_{j+n+1}+\mathrm{h.c.},
\end{equation}
where $c_n$ is a coefficient that depends on the multiple subprocesses involved, e.g., we have $c_2=-1$ and $c_3=-81/64$.
Instead, regardless of $n$ there are always other terms at order one and two that read
\begin{equation}\label{eq:Heff_n}
\begin{aligned}
    \hat{H}^{(1,2)}_{\mathrm{eff},n{=}2} & = -\delta\sum_{j=1}^N\hat{\sigma}^z_j+\frac{h_x^2 n}{4J}\sum_{j=1}^N\Big(\frac{\Pd_{j-1}\hat{\sigma}^z_j\Pd_{j+1}}{n+1}{+}\Pu_{j-1}\hat{\sigma}^z_j\Pd_{j+1}{+}\Pd_{j-1}\hat{\sigma}^z_j\Pu_{j+1}{-}\frac{\Pu_{j-1}\hat{\sigma}^z_j\Pu_{j+1}}{n-1}\Big) \\
    &{+}\frac{h_x^2 n^2}{4J (n-1)}\hspace{-0.1cm}\sum_{j=1}^N\hspace{-0.06cm}\Pu_{j{-}1}\left(\hat{\sigma}^+_j\hat{\sigma}^-_{j{+}1}{+}\hat{\sigma}^-_j\hat{\sigma}^+_{j{+}1}\right)\Pu_{j{+}2}{-}\frac{h_x^2 n^2}{4J (n+1)}\sum_{j=1}^N
    \Pd_{j-1}(\hat{\sigma}^+_j\hat{\sigma}^-_{j+1}+\hat{\sigma}^-_j\hat{\sigma}^+_{j+1})\Pd_{j+2}.
    \end{aligned}
\end{equation}
The terms on the second line create different dynamics. The first one leads to 1-bubbles hopping (as for $n=1$), while the second one allows larger bubbles to exchange down-spins in order to grow or shrink. This allows bubbles of size other than $n$ to develop. This includes the case of larger bubbles shrinking all the way down to 1-bubble and then moving on their own. A sequence of these allowed processes is illustrated in Fig.~\ref{fig:eff} for $n=2$, with higher $n$ values displaying qualitatively similar behaviors. The dynamics at $n>1$ resonances are clearly much richer than at $n=1$. Indeed, while the bubble interaction term should also be present for $n=1$, it cannot act between two 1-bubbles. This would require one of them to shrink to 0, which is not resonant. Thus, in the sector of the $\ket{\uparrow \cdots \uparrow}$ state where only 1-bubbles appear, the bubble interaction term vanishes. 

Finally, it is worth noting that the first term of the effective Hamiltonian at the $n=1$ resonance [Eq.~(\ref{eq:Heff_1})], up to a global spin flip,  is identical to the PXP model used to describe chains of Rydberg atoms~\cite{FendleySachdev,Lesanovsky2012}. The second term can then be recast as $-2\delta\sum_j \Pd_j$ up to an irrelevant constant, and then becomes the chemical potential for the effective PXP model. 
On the other hand, to the best of our knowledge, the effective Hamiltonians for $n>1$ resonances, Eq.~\eqref{eq:Heff_n},  do not map to the models previously studied in the literature.

\section{Acknowledgements}

J.V., D.W. and M.W. acknowledge support from the project Jülich UNified Infrastructure for Quantum computing (JUNIQ) that has received funding from the German Federal Ministry of Education and Research (BMBF) and the Ministry of Culture and Science of the State of North Rhine-Westphalia. A.R. acknowledges support from the project HPCQS (101018180) of the
European High-Performance Computing Joint Undertaking (EuroHPC JU). J.-Y.D., A.H. and Z.P. acknowledge support by the Leverhulme Trust Research Leadership Award RL-2019-015 and EPSRC grants EP/R513258/1, EP/W026848/1. J.-Y.D. acknowledges support from the European Union’s Horizon 2020 research and innovation programme under the Marie Sk\l odowska-Curie Grant Agreement No.101034413. This research was supported in part by grants NSF PHY-1748958 and PHY-2309135 to the Kavli Institute for Theoretical Physics (KITP). Computational portions of this research work were carried out on ARC3 and ARC4, part of the High-Performance Computing facilities at the University of Leeds, UK. G.H. would like to acknowledge the financial support from ARIS, P1-0040 Nonequilibrium Quantum System Dynamics. The authors gratefully acknowledge the Jülich Supercomputing Centre (https://www.fzjuelich.de/ias/jsc) for funding this project by providing computing time on the D-Wave Advantage™ System JUPSI through the Jülich UNified Infrastructure for Quantum computing (JUNIQ). We would like to acknowledge the helpful theoretical discussions with Gianluca Lagnese and the quantum simulation related discussions with D-Wave's experimental team, in particular Allison MacDonald, Gabriel Poulin-Lamarre, Axel Daian and Andrew Berkley. We also thank Victoria Goliber and Andy Mason for patiently organizing and mediating the corresponding meetings that enabled the discussions with D-Wave's team.

\section{Author Contributions Statement}

J.V. conceptualized and performed the quantum simulations. J.V., D.W., F.J. and G.H. designed and analyzed the quantum simulations, whose coherent emulation was performed by A.R. and M.W. 
J-Y.D., A.H. and Z.P. conducted the theoretical analysis. J.V., J-Y.D., D.W. and Z.P. co-wrote the manuscript with input from other authors. All authors participated in the discussions of the results and development of the manuscript.

\section{Competing Interests Statement}

The authors declare no competing interests.

\bibliography{biblio.bib}

\begin{thebibliography}{59}%
\makeatletter
\providecommand \@ifxundefined [1]{%
 \@ifx{#1\undefined}
}%
\providecommand \@ifnum [1]{%
 \ifnum #1\expandafter \@firstoftwo
 \else \expandafter \@secondoftwo
 \fi
}%
\providecommand \@ifx [1]{%
 \ifx #1\expandafter \@firstoftwo
 \else \expandafter \@secondoftwo
 \fi
}%
\providecommand \natexlab [1]{#1}%
\providecommand \enquote  [1]{``#1''}%
\providecommand \bibnamefont  [1]{#1}%
\providecommand \bibfnamefont [1]{#1}%
\providecommand \citenamefont [1]{#1}%
\providecommand \href@noop [0]{\@secondoftwo}%
\providecommand \href [0]{\begingroup \@sanitize@url \@href}%
\providecommand \@href[1]{\@@startlink{#1}\@@href}%
\providecommand \@@href[1]{\endgroup#1\@@endlink}%
\providecommand \@sanitize@url [0]{\catcode `\\12\catcode `\$12\catcode `\&12\catcode `\#12\catcode `\^12\catcode `\_12\catcode `\%12\relax}%
\providecommand \@@startlink[1]{}%
\providecommand \@@endlink[0]{}%
\providecommand \url  [0]{\begingroup\@sanitize@url \@url }%
\providecommand \@url [1]{\endgroup\@href {#1}{\urlprefix }}%
\providecommand \urlprefix  [0]{URL }%
\providecommand \Eprint [0]{\href }%
\providecommand \doibase [0]{https://doi.org/}%
\providecommand \selectlanguage [0]{\@gobble}%
\providecommand \bibinfo  [0]{\@secondoftwo}%
\providecommand \bibfield  [0]{\@secondoftwo}%
\providecommand \translation [1]{[#1]}%
\providecommand \BibitemOpen [0]{}%
\providecommand \bibitemStop [0]{}%
\providecommand \bibitemNoStop [0]{.\EOS\space}%
\providecommand \EOS [0]{\spacefactor3000\relax}%
\providecommand \BibitemShut  [1]{\csname bibitem#1\endcsname}%
\let\auto@bib@innerbib\@empty
\bibitem [{\citenamefont {Coleman}(1977)}]{coleman1977fate}%
  \BibitemOpen
  \bibfield  {author} {\bibinfo {author} {\bibfnamefont {S.}~\bibnamefont {Coleman}},\ }\bibfield  {title} {\bibinfo {title} {Fate of the false vacuum: Semiclassical theory},\ }\href {https://doi.org/10.1103/PhysRevD.15.2929} {\bibfield  {journal} {\bibinfo  {journal} {Phys. Rev. D}\ }\textbf {\bibinfo {volume} {15}},\ \bibinfo {pages} {2929} (\bibinfo {year} {1977})}\BibitemShut {NoStop}%
\bibitem [{\citenamefont {Kobsarev}\ \emph {et~al.}(1974)\citenamefont {Kobsarev}, \citenamefont {Okun},\ and\ \citenamefont {Voloshin}}]{kobsarev1974bubbles}%
  \BibitemOpen
  \bibfield  {author} {\bibinfo {author} {\bibfnamefont {I.}~\bibnamefont {Kobsarev}}, \bibinfo {author} {\bibfnamefont {L.~B.}\ \bibnamefont {Okun}},\ and\ \bibinfo {author} {\bibfnamefont {M.~V.}\ \bibnamefont {Voloshin}},\ }\href {http://inis.iaea.org/search/search.aspx?orig_q=RN:07276129} {\emph {\bibinfo {title} {Bubbles in metastable vacuum}}},\ \bibinfo {type} {Tech. Rep.}\ (\bibinfo  {institution} {Moscow Institute for Theoretical and Experimental Physics},\ \bibinfo {address} {USSR},\ \bibinfo {year} {1974})\ \bibinfo {note} {iTEP--81}\BibitemShut {NoStop}%
\bibitem [{\citenamefont {Linde}(1981)}]{linde1981fate}%
  \BibitemOpen
  \bibfield  {author} {\bibinfo {author} {\bibfnamefont {A.}~\bibnamefont {Linde}},\ }\bibfield  {title} {\bibinfo {title} {Fate of the false vacuum at finite temperature: Theory and applications},\ }\href {https://doi.org/https://doi.org/10.1016/0370-2693(81)90281-1} {\bibfield  {journal} {\bibinfo  {journal} {Physics Letters B}\ }\textbf {\bibinfo {volume} {100}},\ \bibinfo {pages} {37} (\bibinfo {year} {1981})}\BibitemShut {NoStop}%
\bibitem [{\citenamefont {Guth}(1981)}]{guth1981inflationary}%
  \BibitemOpen
  \bibfield  {author} {\bibinfo {author} {\bibfnamefont {A.~H.}\ \bibnamefont {Guth}},\ }\bibfield  {title} {\bibinfo {title} {Inflationary universe: A possible solution to the horizon and flatness problems},\ }\href {https://doi.org/10.1103/PhysRevD.23.347} {\bibfield  {journal} {\bibinfo  {journal} {Phys. Rev. D}\ }\textbf {\bibinfo {volume} {23}},\ \bibinfo {pages} {347} (\bibinfo {year} {1981})}\BibitemShut {NoStop}%
\bibitem [{\citenamefont {Hawking}\ and\ \citenamefont {Moss}(1982)}]{hawking1982supercooled}%
  \BibitemOpen
  \bibfield  {author} {\bibinfo {author} {\bibfnamefont {S.}~\bibnamefont {Hawking}}\ and\ \bibinfo {author} {\bibfnamefont {I.}~\bibnamefont {Moss}},\ }\bibfield  {title} {\bibinfo {title} {Supercooled phase transitions in the very early universe},\ }\href {https://doi.org/https://doi.org/10.1016/0370-2693(82)90946-7} {\bibfield  {journal} {\bibinfo  {journal} {Physics Letters B}\ }\textbf {\bibinfo {volume} {110}},\ \bibinfo {pages} {35} (\bibinfo {year} {1982})}\BibitemShut {NoStop}%
\bibitem [{\citenamefont {Abdalla}\ \emph {et~al.}(2022)\citenamefont {Abdalla}, \citenamefont {Abell{\'a}n}, \citenamefont {Aboubrahim}, \citenamefont {Agnello}, \citenamefont {Akarsu}, \citenamefont {Akrami}, \citenamefont {Alestas}, \citenamefont {Aloni}, \citenamefont {Amendola}, \citenamefont {Anchordoqui} \emph {et~al.}}]{abdalla2022cosmology}%
  \BibitemOpen
  \bibfield  {author} {\bibinfo {author} {\bibfnamefont {E.}~\bibnamefont {Abdalla}}, \bibinfo {author} {\bibfnamefont {G.~F.}\ \bibnamefont {Abell{\'a}n}}, \bibinfo {author} {\bibfnamefont {A.}~\bibnamefont {Aboubrahim}}, \bibinfo {author} {\bibfnamefont {A.}~\bibnamefont {Agnello}}, \bibinfo {author} {\bibfnamefont {{\"O}.}~\bibnamefont {Akarsu}}, \bibinfo {author} {\bibfnamefont {Y.}~\bibnamefont {Akrami}}, \bibinfo {author} {\bibfnamefont {G.}~\bibnamefont {Alestas}}, \bibinfo {author} {\bibfnamefont {D.}~\bibnamefont {Aloni}}, \bibinfo {author} {\bibfnamefont {L.}~\bibnamefont {Amendola}}, \bibinfo {author} {\bibfnamefont {L.~A.}\ \bibnamefont {Anchordoqui}}, \emph {et~al.},\ }\bibfield  {title} {\bibinfo {title} {Cosmology intertwined: A review of the particle physics, astrophysics, and cosmology associated with the cosmological tensions and anomalies},\ }\href {https://doi.org/https://doi.org/10.1016/j.jheap.2022.04.002} {\bibfield  {journal} {\bibinfo  {journal} {Journal of High Energy Astrophysics}\
  }\textbf {\bibinfo {volume} {34}},\ \bibinfo {pages} {49} (\bibinfo {year} {2022})}\BibitemShut {NoStop}%
\bibitem [{\citenamefont {Isidori}\ \emph {et~al.}(2001)\citenamefont {Isidori}, \citenamefont {Ridolfi},\ and\ \citenamefont {Strumia}}]{isidori2001metastability}%
  \BibitemOpen
  \bibfield  {author} {\bibinfo {author} {\bibfnamefont {G.}~\bibnamefont {Isidori}}, \bibinfo {author} {\bibfnamefont {G.}~\bibnamefont {Ridolfi}},\ and\ \bibinfo {author} {\bibfnamefont {A.}~\bibnamefont {Strumia}},\ }\bibfield  {title} {\bibinfo {title} {On the metastability of the {Standard} {Model} vacuum},\ }\href {https://doi.org/https://doi.org/10.1016/S0550-3213(01)00302-9} {\bibfield  {journal} {\bibinfo  {journal} {Nuclear Physics B}\ }\textbf {\bibinfo {volume} {609}},\ \bibinfo {pages} {387} (\bibinfo {year} {2001})}\BibitemShut {NoStop}%
\bibitem [{\citenamefont {Degrassi}\ \emph {et~al.}(2012)\citenamefont {Degrassi}, \citenamefont {Di~Vita}, \citenamefont {Elias-Miro}, \citenamefont {Espinosa}, \citenamefont {Giudice}, \citenamefont {Isidori},\ and\ \citenamefont {Strumia}}]{degrassi2012higgs}%
  \BibitemOpen
  \bibfield  {author} {\bibinfo {author} {\bibfnamefont {G.}~\bibnamefont {Degrassi}}, \bibinfo {author} {\bibfnamefont {S.}~\bibnamefont {Di~Vita}}, \bibinfo {author} {\bibfnamefont {J.}~\bibnamefont {Elias-Miro}}, \bibinfo {author} {\bibfnamefont {J.~R.}\ \bibnamefont {Espinosa}}, \bibinfo {author} {\bibfnamefont {G.~F.}\ \bibnamefont {Giudice}}, \bibinfo {author} {\bibfnamefont {G.}~\bibnamefont {Isidori}},\ and\ \bibinfo {author} {\bibfnamefont {A.}~\bibnamefont {Strumia}},\ }\bibfield  {title} {\bibinfo {title} {Higgs mass and vacuum stability in the standard model at {NNLO}},\ }\href {https://doi.org/10.1007/JHEP08(2012)098} {\bibfield  {journal} {\bibinfo  {journal} {Journal of High Energy Physics}\ }\textbf {\bibinfo {volume} {2012}},\ \bibinfo {pages} {1} (\bibinfo {year} {2012})}\BibitemShut {NoStop}%
\bibitem [{\citenamefont {Caprini}\ \emph {et~al.}(2020)\citenamefont {Caprini}, \citenamefont {Chala}, \citenamefont {Dorsch}, \citenamefont {Hindmarsh}, \citenamefont {Huber}, \citenamefont {Konstandin}, \citenamefont {Kozaczuk}, \citenamefont {Nardini}, \citenamefont {No}, \citenamefont {Rummukainen} \emph {et~al.}}]{caprini2020detecting}%
  \BibitemOpen
  \bibfield  {author} {\bibinfo {author} {\bibfnamefont {C.}~\bibnamefont {Caprini}}, \bibinfo {author} {\bibfnamefont {M.}~\bibnamefont {Chala}}, \bibinfo {author} {\bibfnamefont {G.~C.}\ \bibnamefont {Dorsch}}, \bibinfo {author} {\bibfnamefont {M.}~\bibnamefont {Hindmarsh}}, \bibinfo {author} {\bibfnamefont {S.~J.}\ \bibnamefont {Huber}}, \bibinfo {author} {\bibfnamefont {T.}~\bibnamefont {Konstandin}}, \bibinfo {author} {\bibfnamefont {J.}~\bibnamefont {Kozaczuk}}, \bibinfo {author} {\bibfnamefont {G.}~\bibnamefont {Nardini}}, \bibinfo {author} {\bibfnamefont {J.~M.}\ \bibnamefont {No}}, \bibinfo {author} {\bibfnamefont {K.}~\bibnamefont {Rummukainen}}, \emph {et~al.},\ }\bibfield  {title} {\bibinfo {title} {Detecting gravitational waves from cosmological phase transitions with {LISA}: an update},\ }\href {https://doi.org/10.1088/1475-7516/2020/03/024} {\bibfield  {journal} {\bibinfo  {journal} {Journal of Cosmology and Astroparticle Physics}\ }\textbf {\bibinfo {volume} {2020}}\bibinfo  {number} { (03)},\
  \bibinfo {pages} {024}}\BibitemShut {NoStop}%
\bibitem [{\citenamefont {Farhi}\ \emph {et~al.}(1990)\citenamefont {Farhi}, \citenamefont {Guth},\ and\ \citenamefont {Guven}}]{farhi1990possible}%
  \BibitemOpen
\bibfield  {number} {  }\bibfield  {author} {\bibinfo {author} {\bibfnamefont {E.}~\bibnamefont {Farhi}}, \bibinfo {author} {\bibfnamefont {A.~H.}\ \bibnamefont {Guth}},\ and\ \bibinfo {author} {\bibfnamefont {J.}~\bibnamefont {Guven}},\ }\bibfield  {title} {\bibinfo {title} {Is it possible to create a universe in the laboratory by quantum tunneling?},\ }\href {https://doi.org/10.1016/0550-3213(90)90357-J} {\bibfield  {journal} {\bibinfo  {journal} {Nuclear Physics B}\ }\textbf {\bibinfo {volume} {339}},\ \bibinfo {pages} {417} (\bibinfo {year} {1990})}\BibitemShut {NoStop}%
\bibitem [{\citenamefont {Zurek}(1996)}]{zurek1996cosmological}%
  \BibitemOpen
  \bibfield  {author} {\bibinfo {author} {\bibfnamefont {W.~H.}\ \bibnamefont {Zurek}},\ }\bibfield  {title} {\bibinfo {title} {Cosmological experiments in condensed matter systems},\ }\href {https://doi.org/10.1016/S0370-1573(96)00009-9} {\bibfield  {journal} {\bibinfo  {journal} {Physics Reports}\ }\textbf {\bibinfo {volume} {276}},\ \bibinfo {pages} {177} (\bibinfo {year} {1996})}\BibitemShut {NoStop}%
\bibitem [{\citenamefont {Zenesini}\ \emph {et~al.}(2024)\citenamefont {Zenesini}, \citenamefont {Berti}, \citenamefont {Cominotti}, \citenamefont {Rogora}, \citenamefont {Moss}, \citenamefont {Billam}, \citenamefont {Carusotto}, \citenamefont {Lamporesi}, \citenamefont {Recati},\ and\ \citenamefont {Ferrari}}]{zenesini2024false}%
  \BibitemOpen
  \bibfield  {author} {\bibinfo {author} {\bibfnamefont {A.}~\bibnamefont {Zenesini}}, \bibinfo {author} {\bibfnamefont {A.}~\bibnamefont {Berti}}, \bibinfo {author} {\bibfnamefont {R.}~\bibnamefont {Cominotti}}, \bibinfo {author} {\bibfnamefont {C.}~\bibnamefont {Rogora}}, \bibinfo {author} {\bibfnamefont {I.~G.}\ \bibnamefont {Moss}}, \bibinfo {author} {\bibfnamefont {T.~P.}\ \bibnamefont {Billam}}, \bibinfo {author} {\bibfnamefont {I.}~\bibnamefont {Carusotto}}, \bibinfo {author} {\bibfnamefont {G.}~\bibnamefont {Lamporesi}}, \bibinfo {author} {\bibfnamefont {A.}~\bibnamefont {Recati}},\ and\ \bibinfo {author} {\bibfnamefont {G.}~\bibnamefont {Ferrari}},\ }\bibfield  {title} {\bibinfo {title} {False vacuum decay via bubble formation in ferromagnetic superfluids},\ }\href {https://doi.org/10.1038/s41567-023-02345-4} {\bibfield  {journal} {\bibinfo  {journal} {Nature Physics}\ }\textbf {\bibinfo {volume} {20}},\ \bibinfo {pages} {558} (\bibinfo {year} {2024})}\BibitemShut {NoStop}%
\bibitem [{\citenamefont {Ba{\~{n}}uls}\ \emph {et~al.}(2020)\citenamefont {Ba{\~{n}}uls}, \citenamefont {Blatt}, \citenamefont {Catani}, \citenamefont {Celi}, \citenamefont {Cirac}, \citenamefont {Dalmonte}, \citenamefont {Fallani}, \citenamefont {Jansen}, \citenamefont {Lewenstein}, \citenamefont {Montangero}, \citenamefont {Muschik}, \citenamefont {Reznik}, \citenamefont {Rico}, \citenamefont {Tagliacozzo}, \citenamefont {Van~Acoleyen}, \citenamefont {Verstraete}, \citenamefont {Wiese}, \citenamefont {Wingate}, \citenamefont {Zakrzewski},\ and\ \citenamefont {Zoller}}]{Banuls2020}%
  \BibitemOpen
  \bibfield  {author} {\bibinfo {author} {\bibfnamefont {M.~C.}\ \bibnamefont {Ba{\~{n}}uls}}, \bibinfo {author} {\bibfnamefont {R.}~\bibnamefont {Blatt}}, \bibinfo {author} {\bibfnamefont {J.}~\bibnamefont {Catani}}, \bibinfo {author} {\bibfnamefont {A.}~\bibnamefont {Celi}}, \bibinfo {author} {\bibfnamefont {J.~I.}\ \bibnamefont {Cirac}}, \bibinfo {author} {\bibfnamefont {M.}~\bibnamefont {Dalmonte}}, \bibinfo {author} {\bibfnamefont {L.}~\bibnamefont {Fallani}}, \bibinfo {author} {\bibfnamefont {K.}~\bibnamefont {Jansen}}, \bibinfo {author} {\bibfnamefont {M.}~\bibnamefont {Lewenstein}}, \bibinfo {author} {\bibfnamefont {S.}~\bibnamefont {Montangero}}, \bibinfo {author} {\bibfnamefont {C.~A.}\ \bibnamefont {Muschik}}, \bibinfo {author} {\bibfnamefont {B.}~\bibnamefont {Reznik}}, \bibinfo {author} {\bibfnamefont {E.}~\bibnamefont {Rico}}, \bibinfo {author} {\bibfnamefont {L.}~\bibnamefont {Tagliacozzo}}, \bibinfo {author} {\bibfnamefont {K.}~\bibnamefont {Van~Acoleyen}}, \bibinfo {author} {\bibfnamefont
  {F.}~\bibnamefont {Verstraete}}, \bibinfo {author} {\bibfnamefont {U.-J.}\ \bibnamefont {Wiese}}, \bibinfo {author} {\bibfnamefont {M.}~\bibnamefont {Wingate}}, \bibinfo {author} {\bibfnamefont {J.}~\bibnamefont {Zakrzewski}},\ and\ \bibinfo {author} {\bibfnamefont {P.}~\bibnamefont {Zoller}},\ }\bibfield  {title} {\bibinfo {title} {Simulating lattice gauge theories within quantum technologies},\ }\href {https://doi.org/10.1140/epjd/e2020-100571-8} {\bibfield  {journal} {\bibinfo  {journal} {The European Physical Journal D}\ }\textbf {\bibinfo {volume} {74}},\ \bibinfo {pages} {165} (\bibinfo {year} {2020})}\BibitemShut {NoStop}%
\bibitem [{\citenamefont {Bauer}\ \emph {et~al.}(2023)\citenamefont {Bauer}, \citenamefont {Davoudi}, \citenamefont {Balantekin}, \citenamefont {Bhattacharya}, \citenamefont {Carena}, \citenamefont {de~Jong}, \citenamefont {Draper}, \citenamefont {El-Khadra}, \citenamefont {Gemelke}, \citenamefont {Hanada}, \citenamefont {Kharzeev}, \citenamefont {Lamm}, \citenamefont {Li}, \citenamefont {Liu}, \citenamefont {Lukin}, \citenamefont {Meurice}, \citenamefont {Monroe}, \citenamefont {Nachman}, \citenamefont {Pagano}, \citenamefont {Preskill}, \citenamefont {Rinaldi}, \citenamefont {Roggero}, \citenamefont {Santiago}, \citenamefont {Savage}, \citenamefont {Siddiqi}, \citenamefont {Siopsis}, \citenamefont {Van~Zanten}, \citenamefont {Wiebe}, \citenamefont {Yamauchi}, \citenamefont {Yeter-Aydeniz},\ and\ \citenamefont {Zorzetti}}]{Bauer2023}%
  \BibitemOpen
  \bibfield  {author} {\bibinfo {author} {\bibfnamefont {C.~W.}\ \bibnamefont {Bauer}}, \bibinfo {author} {\bibfnamefont {Z.}~\bibnamefont {Davoudi}}, \bibinfo {author} {\bibfnamefont {A.~B.}\ \bibnamefont {Balantekin}}, \bibinfo {author} {\bibfnamefont {T.}~\bibnamefont {Bhattacharya}}, \bibinfo {author} {\bibfnamefont {M.}~\bibnamefont {Carena}}, \bibinfo {author} {\bibfnamefont {W.~A.}\ \bibnamefont {de~Jong}}, \bibinfo {author} {\bibfnamefont {P.}~\bibnamefont {Draper}}, \bibinfo {author} {\bibfnamefont {A.}~\bibnamefont {El-Khadra}}, \bibinfo {author} {\bibfnamefont {N.}~\bibnamefont {Gemelke}}, \bibinfo {author} {\bibfnamefont {M.}~\bibnamefont {Hanada}}, \bibinfo {author} {\bibfnamefont {D.}~\bibnamefont {Kharzeev}}, \bibinfo {author} {\bibfnamefont {H.}~\bibnamefont {Lamm}}, \bibinfo {author} {\bibfnamefont {Y.-Y.}\ \bibnamefont {Li}}, \bibinfo {author} {\bibfnamefont {J.}~\bibnamefont {Liu}}, \bibinfo {author} {\bibfnamefont {M.}~\bibnamefont {Lukin}}, \bibinfo {author} {\bibfnamefont
  {Y.}~\bibnamefont {Meurice}}, \bibinfo {author} {\bibfnamefont {C.}~\bibnamefont {Monroe}}, \bibinfo {author} {\bibfnamefont {B.}~\bibnamefont {Nachman}}, \bibinfo {author} {\bibfnamefont {G.}~\bibnamefont {Pagano}}, \bibinfo {author} {\bibfnamefont {J.}~\bibnamefont {Preskill}}, \bibinfo {author} {\bibfnamefont {E.}~\bibnamefont {Rinaldi}}, \bibinfo {author} {\bibfnamefont {A.}~\bibnamefont {Roggero}}, \bibinfo {author} {\bibfnamefont {D.~I.}\ \bibnamefont {Santiago}}, \bibinfo {author} {\bibfnamefont {M.~J.}\ \bibnamefont {Savage}}, \bibinfo {author} {\bibfnamefont {I.}~\bibnamefont {Siddiqi}}, \bibinfo {author} {\bibfnamefont {G.}~\bibnamefont {Siopsis}}, \bibinfo {author} {\bibfnamefont {D.}~\bibnamefont {Van~Zanten}}, \bibinfo {author} {\bibfnamefont {N.}~\bibnamefont {Wiebe}}, \bibinfo {author} {\bibfnamefont {Y.}~\bibnamefont {Yamauchi}}, \bibinfo {author} {\bibfnamefont {K.}~\bibnamefont {Yeter-Aydeniz}},\ and\ \bibinfo {author} {\bibfnamefont {S.}~\bibnamefont {Zorzetti}},\ }\bibfield  {title}
  {\bibinfo {title} {Quantum simulation for high-energy physics},\ }\href {https://doi.org/10.1103/PRXQuantum.4.027001} {\bibfield  {journal} {\bibinfo  {journal} {PRX Quantum}\ }\textbf {\bibinfo {volume} {4}},\ \bibinfo {pages} {027001} (\bibinfo {year} {2023})}\BibitemShut {NoStop}%
\bibitem [{\citenamefont {Halimeh}\ \emph {et~al.}(2023)\citenamefont {Halimeh}, \citenamefont {Aidelsburger}, \citenamefont {Grusdt}, \citenamefont {Hauke},\ and\ \citenamefont {Yang}}]{halimeh2023coldatom}%
  \BibitemOpen
  \bibfield  {author} {\bibinfo {author} {\bibfnamefont {J.~C.}\ \bibnamefont {Halimeh}}, \bibinfo {author} {\bibfnamefont {M.}~\bibnamefont {Aidelsburger}}, \bibinfo {author} {\bibfnamefont {F.}~\bibnamefont {Grusdt}}, \bibinfo {author} {\bibfnamefont {P.}~\bibnamefont {Hauke}},\ and\ \bibinfo {author} {\bibfnamefont {B.}~\bibnamefont {Yang}},\ }\href@noop {} {\bibinfo {title} {Cold-atom quantum simulators of gauge theories}} (\bibinfo {year} {2023}),\ \Eprint {https://arxiv.org/abs/2310.12201} {arXiv:2310.12201 [cond-mat.quant-gas]} \BibitemShut {NoStop}%
\bibitem [{\citenamefont {Billam}\ \emph {et~al.}(2019)\citenamefont {Billam}, \citenamefont {Gregory}, \citenamefont {Michel},\ and\ \citenamefont {Moss}}]{billam2019simulating}%
  \BibitemOpen
  \bibfield  {author} {\bibinfo {author} {\bibfnamefont {T.~P.}\ \bibnamefont {Billam}}, \bibinfo {author} {\bibfnamefont {R.}~\bibnamefont {Gregory}}, \bibinfo {author} {\bibfnamefont {F.}~\bibnamefont {Michel}},\ and\ \bibinfo {author} {\bibfnamefont {I.~G.}\ \bibnamefont {Moss}},\ }\bibfield  {title} {\bibinfo {title} {Simulating seeded vacuum decay in a cold atom system},\ }\href {https://doi.org/10.1103/PhysRevD.100.065016} {\bibfield  {journal} {\bibinfo  {journal} {Phys. Rev. D}\ }\textbf {\bibinfo {volume} {100}},\ \bibinfo {pages} {065016} (\bibinfo {year} {2019})}\BibitemShut {NoStop}%
\bibitem [{\citenamefont {Billam}\ \emph {et~al.}(2020)\citenamefont {Billam}, \citenamefont {Brown},\ and\ \citenamefont {Moss}}]{billam2020simulating}%
  \BibitemOpen
  \bibfield  {author} {\bibinfo {author} {\bibfnamefont {T.~P.}\ \bibnamefont {Billam}}, \bibinfo {author} {\bibfnamefont {K.}~\bibnamefont {Brown}},\ and\ \bibinfo {author} {\bibfnamefont {I.~G.}\ \bibnamefont {Moss}},\ }\bibfield  {title} {\bibinfo {title} {Simulating cosmological supercooling with a cold-atom system},\ }\href {https://doi.org/10.1103/PhysRevA.102.043324} {\bibfield  {journal} {\bibinfo  {journal} {Phys. Rev. A}\ }\textbf {\bibinfo {volume} {102}},\ \bibinfo {pages} {043324} (\bibinfo {year} {2020})}\BibitemShut {NoStop}%
\bibitem [{\citenamefont {Abel}\ and\ \citenamefont {Spannowsky}(2021)}]{abel2021quantum}%
  \BibitemOpen
  \bibfield  {author} {\bibinfo {author} {\bibfnamefont {S.}~\bibnamefont {Abel}}\ and\ \bibinfo {author} {\bibfnamefont {M.}~\bibnamefont {Spannowsky}},\ }\bibfield  {title} {\bibinfo {title} {Quantum-field-theoretic simulation platform for observing the fate of the false vacuum},\ }\href {https://doi.org/10.1103/PRXQuantum.2.010349} {\bibfield  {journal} {\bibinfo  {journal} {PRX Quantum}\ }\textbf {\bibinfo {volume} {2}},\ \bibinfo {pages} {010349} (\bibinfo {year} {2021})}\BibitemShut {NoStop}%
\bibitem [{\citenamefont {Ng}\ \emph {et~al.}(2021)\citenamefont {Ng}, \citenamefont {Opanchuk}, \citenamefont {Thenabadu}, \citenamefont {Reid},\ and\ \citenamefont {Drummond}}]{ng2021fate}%
  \BibitemOpen
  \bibfield  {author} {\bibinfo {author} {\bibfnamefont {K.~L.}\ \bibnamefont {Ng}}, \bibinfo {author} {\bibfnamefont {B.}~\bibnamefont {Opanchuk}}, \bibinfo {author} {\bibfnamefont {M.}~\bibnamefont {Thenabadu}}, \bibinfo {author} {\bibfnamefont {M.}~\bibnamefont {Reid}},\ and\ \bibinfo {author} {\bibfnamefont {P.~D.}\ \bibnamefont {Drummond}},\ }\bibfield  {title} {\bibinfo {title} {Fate of the false vacuum: Finite temperature, entropy, and topological phase in quantum simulations of the early universe},\ }\href {https://doi.org/10.1103/PRXQuantum.2.010350} {\bibfield  {journal} {\bibinfo  {journal} {PRX Quantum}\ }\textbf {\bibinfo {volume} {2}},\ \bibinfo {pages} {010350} (\bibinfo {year} {2021})}\BibitemShut {NoStop}%
\bibitem [{\citenamefont {Milsted}\ \emph {et~al.}(2022)\citenamefont {Milsted}, \citenamefont {Liu}, \citenamefont {Preskill},\ and\ \citenamefont {Vidal}}]{milsted2022collisions}%
  \BibitemOpen
  \bibfield  {author} {\bibinfo {author} {\bibfnamefont {A.}~\bibnamefont {Milsted}}, \bibinfo {author} {\bibfnamefont {J.}~\bibnamefont {Liu}}, \bibinfo {author} {\bibfnamefont {J.}~\bibnamefont {Preskill}},\ and\ \bibinfo {author} {\bibfnamefont {G.}~\bibnamefont {Vidal}},\ }\bibfield  {title} {\bibinfo {title} {Collisions of false-vacuum bubble walls in a quantum spin chain},\ }\href {https://doi.org/10.1103/PRXQuantum.3.020316} {\bibfield  {journal} {\bibinfo  {journal} {PRX Quantum}\ }\textbf {\bibinfo {volume} {3}},\ \bibinfo {pages} {020316} (\bibinfo {year} {2022})}\BibitemShut {NoStop}%
\bibitem [{\citenamefont {Darbha}\ \emph {et~al.}(2024{\natexlab{a}})\citenamefont {Darbha}, \citenamefont {Kornjača}, \citenamefont {Liu}, \citenamefont {Balewski}, \citenamefont {Hirsbrunner}, \citenamefont {Lopes}, \citenamefont {Wang}, \citenamefont {Beeumen}, \citenamefont {Camps},\ and\ \citenamefont {Klymko}}]{Darbha2024_1}%
  \BibitemOpen
  \bibfield  {author} {\bibinfo {author} {\bibfnamefont {S.}~\bibnamefont {Darbha}}, \bibinfo {author} {\bibfnamefont {M.}~\bibnamefont {Kornjača}}, \bibinfo {author} {\bibfnamefont {F.}~\bibnamefont {Liu}}, \bibinfo {author} {\bibfnamefont {J.}~\bibnamefont {Balewski}}, \bibinfo {author} {\bibfnamefont {M.~R.}\ \bibnamefont {Hirsbrunner}}, \bibinfo {author} {\bibfnamefont {P.}~\bibnamefont {Lopes}}, \bibinfo {author} {\bibfnamefont {S.-T.}\ \bibnamefont {Wang}}, \bibinfo {author} {\bibfnamefont {R.~V.}\ \bibnamefont {Beeumen}}, \bibinfo {author} {\bibfnamefont {D.}~\bibnamefont {Camps}},\ and\ \bibinfo {author} {\bibfnamefont {K.}~\bibnamefont {Klymko}},\ }\href@noop {} {\bibinfo {title} {False vacuum decay and nucleation dynamics in neutral atom systems}} (\bibinfo {year} {2024}{\natexlab{a}}),\ \Eprint {https://arxiv.org/abs/2404.12360} {arXiv:2404.12360 [quant-ph]} \BibitemShut {NoStop}%
\bibitem [{\citenamefont {Darbha}\ \emph {et~al.}(2024{\natexlab{b}})\citenamefont {Darbha}, \citenamefont {Kornjača}, \citenamefont {Liu}, \citenamefont {Balewski}, \citenamefont {Hirsbrunner}, \citenamefont {Lopes}, \citenamefont {Wang}, \citenamefont {Beeumen}, \citenamefont {Klymko},\ and\ \citenamefont {Camps}}]{Darbha2024_2}%
  \BibitemOpen
  \bibfield  {author} {\bibinfo {author} {\bibfnamefont {S.}~\bibnamefont {Darbha}}, \bibinfo {author} {\bibfnamefont {M.}~\bibnamefont {Kornjača}}, \bibinfo {author} {\bibfnamefont {F.}~\bibnamefont {Liu}}, \bibinfo {author} {\bibfnamefont {J.}~\bibnamefont {Balewski}}, \bibinfo {author} {\bibfnamefont {M.~R.}\ \bibnamefont {Hirsbrunner}}, \bibinfo {author} {\bibfnamefont {P.}~\bibnamefont {Lopes}}, \bibinfo {author} {\bibfnamefont {S.-T.}\ \bibnamefont {Wang}}, \bibinfo {author} {\bibfnamefont {R.~V.}\ \bibnamefont {Beeumen}}, \bibinfo {author} {\bibfnamefont {K.}~\bibnamefont {Klymko}},\ and\ \bibinfo {author} {\bibfnamefont {D.}~\bibnamefont {Camps}},\ }\href@noop {} {\bibinfo {title} {Long-lived oscillations of false and true vacuum states in neutral atom systems}} (\bibinfo {year} {2024}{\natexlab{b}}),\ \Eprint {https://arxiv.org/abs/2404.12371} {arXiv:2404.12371 [quant-ph]} \BibitemShut {NoStop}%
\bibitem [{\citenamefont {Harris}\ \emph {et~al.}(2018)\citenamefont {Harris}, \citenamefont {Sato}, \citenamefont {Berkley}, \citenamefont {Reis}, \citenamefont {Altomare}, \citenamefont {Amin}, \citenamefont {Boothby}, \citenamefont {Bunyk}, \citenamefont {Deng}, \citenamefont {Enderud}, \citenamefont {Huang}, \citenamefont {Hoskinson}, \citenamefont {Johnson}, \citenamefont {Ladizinsky}, \citenamefont {Ladizinsky}, \citenamefont {Lanting}, \citenamefont {Li}, \citenamefont {Medina}, \citenamefont {Molavi}, \citenamefont {Neufeld}, \citenamefont {Oh}, \citenamefont {Pavlov}, \citenamefont {Perminov}, \citenamefont {Poulin-Lamarre}, \citenamefont {Rich}, \citenamefont {Smirnov}, \citenamefont {Swenson}, \citenamefont {Tsai}, \citenamefont {Volkmann}, \citenamefont {Whittaker},\ and\ \citenamefont {Yao}}]{harris2018phase}%
  \BibitemOpen
  \bibfield  {author} {\bibinfo {author} {\bibfnamefont {R.}~\bibnamefont {Harris}}, \bibinfo {author} {\bibfnamefont {Y.}~\bibnamefont {Sato}}, \bibinfo {author} {\bibfnamefont {A.~J.}\ \bibnamefont {Berkley}}, \bibinfo {author} {\bibfnamefont {M.}~\bibnamefont {Reis}}, \bibinfo {author} {\bibfnamefont {F.}~\bibnamefont {Altomare}}, \bibinfo {author} {\bibfnamefont {M.~H.}\ \bibnamefont {Amin}}, \bibinfo {author} {\bibfnamefont {K.}~\bibnamefont {Boothby}}, \bibinfo {author} {\bibfnamefont {P.}~\bibnamefont {Bunyk}}, \bibinfo {author} {\bibfnamefont {C.}~\bibnamefont {Deng}}, \bibinfo {author} {\bibfnamefont {C.}~\bibnamefont {Enderud}}, \bibinfo {author} {\bibfnamefont {S.}~\bibnamefont {Huang}}, \bibinfo {author} {\bibfnamefont {E.}~\bibnamefont {Hoskinson}}, \bibinfo {author} {\bibfnamefont {M.~W.}\ \bibnamefont {Johnson}}, \bibinfo {author} {\bibfnamefont {E.}~\bibnamefont {Ladizinsky}}, \bibinfo {author} {\bibfnamefont {N.}~\bibnamefont {Ladizinsky}}, \bibinfo {author} {\bibfnamefont {T.}~\bibnamefont
  {Lanting}}, \bibinfo {author} {\bibfnamefont {R.}~\bibnamefont {Li}}, \bibinfo {author} {\bibfnamefont {T.}~\bibnamefont {Medina}}, \bibinfo {author} {\bibfnamefont {R.}~\bibnamefont {Molavi}}, \bibinfo {author} {\bibfnamefont {R.}~\bibnamefont {Neufeld}}, \bibinfo {author} {\bibfnamefont {T.}~\bibnamefont {Oh}}, \bibinfo {author} {\bibfnamefont {I.}~\bibnamefont {Pavlov}}, \bibinfo {author} {\bibfnamefont {I.}~\bibnamefont {Perminov}}, \bibinfo {author} {\bibfnamefont {G.}~\bibnamefont {Poulin-Lamarre}}, \bibinfo {author} {\bibfnamefont {C.}~\bibnamefont {Rich}}, \bibinfo {author} {\bibfnamefont {A.}~\bibnamefont {Smirnov}}, \bibinfo {author} {\bibfnamefont {L.}~\bibnamefont {Swenson}}, \bibinfo {author} {\bibfnamefont {N.}~\bibnamefont {Tsai}}, \bibinfo {author} {\bibfnamefont {M.}~\bibnamefont {Volkmann}}, \bibinfo {author} {\bibfnamefont {J.}~\bibnamefont {Whittaker}},\ and\ \bibinfo {author} {\bibfnamefont {J.}~\bibnamefont {Yao}},\ }\bibfield  {title} {\bibinfo {title} {Phase transitions in a
  programmable quantum spin glass simulator},\ }\href {https://doi.org/10.1126/science.aat2025} {\bibfield  {journal} {\bibinfo  {journal} {Science}\ }\textbf {\bibinfo {volume} {361}},\ \bibinfo {pages} {162} (\bibinfo {year} {2018})}\BibitemShut {NoStop}%
\bibitem [{\citenamefont {Bando}\ \emph {et~al.}(2020)\citenamefont {Bando}, \citenamefont {Susa}, \citenamefont {Oshiyama}, \citenamefont {Shibata}, \citenamefont {Ohzeki}, \citenamefont {G\'omez-Ruiz}, \citenamefont {Lidar}, \citenamefont {Suzuki}, \citenamefont {del Campo},\ and\ \citenamefont {Nishimori}}]{bando2020probing}%
  \BibitemOpen
  \bibfield  {author} {\bibinfo {author} {\bibfnamefont {Y.}~\bibnamefont {Bando}}, \bibinfo {author} {\bibfnamefont {Y.}~\bibnamefont {Susa}}, \bibinfo {author} {\bibfnamefont {H.}~\bibnamefont {Oshiyama}}, \bibinfo {author} {\bibfnamefont {N.}~\bibnamefont {Shibata}}, \bibinfo {author} {\bibfnamefont {M.}~\bibnamefont {Ohzeki}}, \bibinfo {author} {\bibfnamefont {F.~J.}\ \bibnamefont {G\'omez-Ruiz}}, \bibinfo {author} {\bibfnamefont {D.~A.}\ \bibnamefont {Lidar}}, \bibinfo {author} {\bibfnamefont {S.}~\bibnamefont {Suzuki}}, \bibinfo {author} {\bibfnamefont {A.}~\bibnamefont {del Campo}},\ and\ \bibinfo {author} {\bibfnamefont {H.}~\bibnamefont {Nishimori}},\ }\bibfield  {title} {\bibinfo {title} {Probing the universality of topological defect formation in a quantum annealer: Kibble-zurek mechanism and beyond},\ }\href {https://doi.org/10.1103/PhysRevResearch.2.033369} {\bibfield  {journal} {\bibinfo  {journal} {Phys. Rev. Res.}\ }\textbf {\bibinfo {volume} {2}},\ \bibinfo {pages} {033369} (\bibinfo {year}
  {2020})}\BibitemShut {NoStop}%
\bibitem [{\citenamefont {King}\ \emph {et~al.}(2022)\citenamefont {King}, \citenamefont {Suzuki}, \citenamefont {Raymond}, \citenamefont {Zucca}, \citenamefont {Lanting}, \citenamefont {Altomare}, \citenamefont {Berkley}, \citenamefont {Ejtemaee}, \citenamefont {Hoskinson}, \citenamefont {Huang}, \citenamefont {Ladizinsky}, \citenamefont {MacDonald}, \citenamefont {Marsden}, \citenamefont {Oh}, \citenamefont {Poulin-Lamarre}, \citenamefont {Reis}, \citenamefont {Rich}, \citenamefont {Sato}, \citenamefont {Whittaker}, \citenamefont {Yao}, \citenamefont {Harris}, \citenamefont {Lidar}, \citenamefont {Nishimori},\ and\ \citenamefont {Amin}}]{king2022coherent}%
  \BibitemOpen
  \bibfield  {author} {\bibinfo {author} {\bibfnamefont {A.~D.}\ \bibnamefont {King}}, \bibinfo {author} {\bibfnamefont {S.}~\bibnamefont {Suzuki}}, \bibinfo {author} {\bibfnamefont {J.}~\bibnamefont {Raymond}}, \bibinfo {author} {\bibfnamefont {A.}~\bibnamefont {Zucca}}, \bibinfo {author} {\bibfnamefont {T.}~\bibnamefont {Lanting}}, \bibinfo {author} {\bibfnamefont {F.}~\bibnamefont {Altomare}}, \bibinfo {author} {\bibfnamefont {A.~J.}\ \bibnamefont {Berkley}}, \bibinfo {author} {\bibfnamefont {S.}~\bibnamefont {Ejtemaee}}, \bibinfo {author} {\bibfnamefont {E.}~\bibnamefont {Hoskinson}}, \bibinfo {author} {\bibfnamefont {S.}~\bibnamefont {Huang}}, \bibinfo {author} {\bibfnamefont {E.}~\bibnamefont {Ladizinsky}}, \bibinfo {author} {\bibfnamefont {A.~J.~R.}\ \bibnamefont {MacDonald}}, \bibinfo {author} {\bibfnamefont {G.}~\bibnamefont {Marsden}}, \bibinfo {author} {\bibfnamefont {T.}~\bibnamefont {Oh}}, \bibinfo {author} {\bibfnamefont {G.}~\bibnamefont {Poulin-Lamarre}}, \bibinfo {author} {\bibfnamefont
  {M.}~\bibnamefont {Reis}}, \bibinfo {author} {\bibfnamefont {C.}~\bibnamefont {Rich}}, \bibinfo {author} {\bibfnamefont {Y.}~\bibnamefont {Sato}}, \bibinfo {author} {\bibfnamefont {J.~D.}\ \bibnamefont {Whittaker}}, \bibinfo {author} {\bibfnamefont {J.}~\bibnamefont {Yao}}, \bibinfo {author} {\bibfnamefont {R.}~\bibnamefont {Harris}}, \bibinfo {author} {\bibfnamefont {D.~A.}\ \bibnamefont {Lidar}}, \bibinfo {author} {\bibfnamefont {H.}~\bibnamefont {Nishimori}},\ and\ \bibinfo {author} {\bibfnamefont {M.~H.}\ \bibnamefont {Amin}},\ }\bibfield  {title} {\bibinfo {title} {Coherent quantum annealing in a programmable 2,000{\thinspace}qubit ising chain},\ }\href {https://doi.org/10.1038/s41567-022-01741-6} {\bibfield  {journal} {\bibinfo  {journal} {Nature Physics}\ }\textbf {\bibinfo {volume} {18}},\ \bibinfo {pages} {1324} (\bibinfo {year} {2022})}\BibitemShut {NoStop}%
\bibitem [{\citenamefont {King}\ \emph {et~al.}(2023)\citenamefont {King}, \citenamefont {Raymond}, \citenamefont {Lanting}, \citenamefont {Harris}, \citenamefont {Zucca}, \citenamefont {Altomare}, \citenamefont {Berkley}, \citenamefont {Boothby}, \citenamefont {Ejtemaee}, \citenamefont {Enderud}, \citenamefont {Hoskinson}, \citenamefont {Huang}, \citenamefont {Ladizinsky}, \citenamefont {MacDonald}, \citenamefont {Marsden}, \citenamefont {Molavi}, \citenamefont {Oh}, \citenamefont {Poulin-Lamarre}, \citenamefont {Reis}, \citenamefont {Rich}, \citenamefont {Sato}, \citenamefont {Tsai}, \citenamefont {Volkmann}, \citenamefont {Whittaker}, \citenamefont {Yao}, \citenamefont {Sandvik},\ and\ \citenamefont {Amin}}]{king2023quantum}%
  \BibitemOpen
  \bibfield  {author} {\bibinfo {author} {\bibfnamefont {A.~D.}\ \bibnamefont {King}}, \bibinfo {author} {\bibfnamefont {J.}~\bibnamefont {Raymond}}, \bibinfo {author} {\bibfnamefont {T.}~\bibnamefont {Lanting}}, \bibinfo {author} {\bibfnamefont {R.}~\bibnamefont {Harris}}, \bibinfo {author} {\bibfnamefont {A.}~\bibnamefont {Zucca}}, \bibinfo {author} {\bibfnamefont {F.}~\bibnamefont {Altomare}}, \bibinfo {author} {\bibfnamefont {A.~J.}\ \bibnamefont {Berkley}}, \bibinfo {author} {\bibfnamefont {K.}~\bibnamefont {Boothby}}, \bibinfo {author} {\bibfnamefont {S.}~\bibnamefont {Ejtemaee}}, \bibinfo {author} {\bibfnamefont {C.}~\bibnamefont {Enderud}}, \bibinfo {author} {\bibfnamefont {E.}~\bibnamefont {Hoskinson}}, \bibinfo {author} {\bibfnamefont {S.}~\bibnamefont {Huang}}, \bibinfo {author} {\bibfnamefont {E.}~\bibnamefont {Ladizinsky}}, \bibinfo {author} {\bibfnamefont {A.~J.~R.}\ \bibnamefont {MacDonald}}, \bibinfo {author} {\bibfnamefont {G.}~\bibnamefont {Marsden}}, \bibinfo {author} {\bibfnamefont
  {R.}~\bibnamefont {Molavi}}, \bibinfo {author} {\bibfnamefont {T.}~\bibnamefont {Oh}}, \bibinfo {author} {\bibfnamefont {G.}~\bibnamefont {Poulin-Lamarre}}, \bibinfo {author} {\bibfnamefont {M.}~\bibnamefont {Reis}}, \bibinfo {author} {\bibfnamefont {C.}~\bibnamefont {Rich}}, \bibinfo {author} {\bibfnamefont {Y.}~\bibnamefont {Sato}}, \bibinfo {author} {\bibfnamefont {N.}~\bibnamefont {Tsai}}, \bibinfo {author} {\bibfnamefont {M.}~\bibnamefont {Volkmann}}, \bibinfo {author} {\bibfnamefont {J.~D.}\ \bibnamefont {Whittaker}}, \bibinfo {author} {\bibfnamefont {J.}~\bibnamefont {Yao}}, \bibinfo {author} {\bibfnamefont {A.~W.}\ \bibnamefont {Sandvik}},\ and\ \bibinfo {author} {\bibfnamefont {M.~H.}\ \bibnamefont {Amin}},\ }\bibfield  {title} {\bibinfo {title} {Quantum critical dynamics in a 5,000-qubit programmable spin glass},\ }\href {https://doi.org/10.1038/s41586-023-05867-2} {\bibfield  {journal} {\bibinfo  {journal} {Nature}\ }\textbf {\bibinfo {volume} {617}},\ \bibinfo {pages} {61} (\bibinfo {year}
  {2023})}\BibitemShut {NoStop}%
\bibitem [{\citenamefont {Kormos}\ \emph {et~al.}(2017)\citenamefont {Kormos}, \citenamefont {Collura}, \citenamefont {Tak{\'a}cs},\ and\ \citenamefont {Calabrese}}]{kormos2017real}%
  \BibitemOpen
  \bibfield  {author} {\bibinfo {author} {\bibfnamefont {M.}~\bibnamefont {Kormos}}, \bibinfo {author} {\bibfnamefont {M.}~\bibnamefont {Collura}}, \bibinfo {author} {\bibfnamefont {G.}~\bibnamefont {Tak{\'a}cs}},\ and\ \bibinfo {author} {\bibfnamefont {P.}~\bibnamefont {Calabrese}},\ }\bibfield  {title} {\bibinfo {title} {Real-time confinement following a quantum quench to a non-integrable model},\ }\href {https://doi.org/10.1038/nphys3934} {\bibfield  {journal} {\bibinfo  {journal} {Nat. Phys.}\ }\textbf {\bibinfo {volume} {13}},\ \bibinfo {pages} {246} (\bibinfo {year} {2017})}\BibitemShut {NoStop}%
\bibitem [{\citenamefont {Liu}\ \emph {et~al.}(2019)\citenamefont {Liu}, \citenamefont {Lundgren}, \citenamefont {Titum}, \citenamefont {Pagano}, \citenamefont {Zhang}, \citenamefont {Monroe},\ and\ \citenamefont {Gorshkov}}]{liu2019confined}%
  \BibitemOpen
  \bibfield  {author} {\bibinfo {author} {\bibfnamefont {F.}~\bibnamefont {Liu}}, \bibinfo {author} {\bibfnamefont {R.}~\bibnamefont {Lundgren}}, \bibinfo {author} {\bibfnamefont {P.}~\bibnamefont {Titum}}, \bibinfo {author} {\bibfnamefont {G.}~\bibnamefont {Pagano}}, \bibinfo {author} {\bibfnamefont {J.}~\bibnamefont {Zhang}}, \bibinfo {author} {\bibfnamefont {C.}~\bibnamefont {Monroe}},\ and\ \bibinfo {author} {\bibfnamefont {A.~V.}\ \bibnamefont {Gorshkov}},\ }\bibfield  {title} {\bibinfo {title} {Confined quasiparticle dynamics in long-range interacting quantum spin chains},\ }\href {https://doi.org/10.1103/PhysRevLett.122.150601} {\bibfield  {journal} {\bibinfo  {journal} {Phys. Rev. Lett.}\ }\textbf {\bibinfo {volume} {122}},\ \bibinfo {pages} {150601} (\bibinfo {year} {2019})}\BibitemShut {NoStop}%
\bibitem [{\citenamefont {Tan}\ \emph {et~al.}(2021)\citenamefont {Tan}, \citenamefont {Becker}, \citenamefont {Liu}, \citenamefont {Pagano}, \citenamefont {Collins}, \citenamefont {De}, \citenamefont {Feng}, \citenamefont {Kaplan}, \citenamefont {Kyprianidis}, \citenamefont {Lundgren} \emph {et~al.}}]{tan2021domain}%
  \BibitemOpen
  \bibfield  {author} {\bibinfo {author} {\bibfnamefont {W.~L.}\ \bibnamefont {Tan}}, \bibinfo {author} {\bibfnamefont {P.}~\bibnamefont {Becker}}, \bibinfo {author} {\bibfnamefont {F.}~\bibnamefont {Liu}}, \bibinfo {author} {\bibfnamefont {G.}~\bibnamefont {Pagano}}, \bibinfo {author} {\bibfnamefont {K.}~\bibnamefont {Collins}}, \bibinfo {author} {\bibfnamefont {A.}~\bibnamefont {De}}, \bibinfo {author} {\bibfnamefont {L.}~\bibnamefont {Feng}}, \bibinfo {author} {\bibfnamefont {H.}~\bibnamefont {Kaplan}}, \bibinfo {author} {\bibfnamefont {A.}~\bibnamefont {Kyprianidis}}, \bibinfo {author} {\bibfnamefont {R.}~\bibnamefont {Lundgren}}, \emph {et~al.},\ }\bibfield  {title} {\bibinfo {title} {Domain-wall confinement and dynamics in a quantum simulator},\ }\href {https://doi.org/10.1038/s41567-021-01194-3} {\bibfield  {journal} {\bibinfo  {journal} {Nat. Phys.}\ }\textbf {\bibinfo {volume} {17}},\ \bibinfo {pages} {742} (\bibinfo {year} {2021})}\BibitemShut {NoStop}%
\bibitem [{\citenamefont {Vovrosh}\ and\ \citenamefont {Knolle}(2021)}]{vovrosh2021confinement}%
  \BibitemOpen
  \bibfield  {author} {\bibinfo {author} {\bibfnamefont {J.}~\bibnamefont {Vovrosh}}\ and\ \bibinfo {author} {\bibfnamefont {J.}~\bibnamefont {Knolle}},\ }\bibfield  {title} {\bibinfo {title} {Confinement and entanglement dynamics on a digital quantum computer},\ }\href {https://doi.org/10.1038/s41598-021-90849-5} {\bibfield  {journal} {\bibinfo  {journal} {Scientific reports}\ }\textbf {\bibinfo {volume} {11}},\ \bibinfo {pages} {11577} (\bibinfo {year} {2021})}\BibitemShut {NoStop}%
\bibitem [{\citenamefont {Lagnese}\ \emph {et~al.}(2023)\citenamefont {Lagnese}, \citenamefont {Surace}, \citenamefont {Morampudi},\ and\ \citenamefont {Wilczek}}]{Lagnese2023}%
  \BibitemOpen
  \bibfield  {author} {\bibinfo {author} {\bibfnamefont {G.}~\bibnamefont {Lagnese}}, \bibinfo {author} {\bibfnamefont {F.~M.}\ \bibnamefont {Surace}}, \bibinfo {author} {\bibfnamefont {S.}~\bibnamefont {Morampudi}},\ and\ \bibinfo {author} {\bibfnamefont {F.}~\bibnamefont {Wilczek}},\ }\href@noop {} {\bibinfo {title} {Detecting a long lived false vacuum with quantum quenches}} (\bibinfo {year} {2023}),\ \Eprint {https://arxiv.org/abs/2308.08340} {arXiv:2308.08340 [cond-mat.stat-mech]} \BibitemShut {NoStop}%
\bibitem [{\citenamefont {Rutkevich}(1999)}]{rutkevich1999decay}%
  \BibitemOpen
  \bibfield  {author} {\bibinfo {author} {\bibfnamefont {S.~B.}\ \bibnamefont {Rutkevich}},\ }\bibfield  {title} {\bibinfo {title} {Decay of the metastable phase in $d=1$ and $d=2$ ising models},\ }\href {https://doi.org/10.1103/PhysRevB.60.14525} {\bibfield  {journal} {\bibinfo  {journal} {Phys. Rev. B}\ }\textbf {\bibinfo {volume} {60}},\ \bibinfo {pages} {14525} (\bibinfo {year} {1999})}\BibitemShut {NoStop}%
\bibitem [{\citenamefont {Lagnese}\ \emph {et~al.}(2021)\citenamefont {Lagnese}, \citenamefont {Surace}, \citenamefont {Kormos},\ and\ \citenamefont {Calabrese}}]{lagnese2021false}%
  \BibitemOpen
  \bibfield  {author} {\bibinfo {author} {\bibfnamefont {G.}~\bibnamefont {Lagnese}}, \bibinfo {author} {\bibfnamefont {F.~M.}\ \bibnamefont {Surace}}, \bibinfo {author} {\bibfnamefont {M.}~\bibnamefont {Kormos}},\ and\ \bibinfo {author} {\bibfnamefont {P.}~\bibnamefont {Calabrese}},\ }\bibfield  {title} {\bibinfo {title} {False vacuum decay in quantum spin chains},\ }\href {https://doi.org/10.1103/PhysRevB.104.L201106} {\bibfield  {journal} {\bibinfo  {journal} {Phys. Rev. B}\ }\textbf {\bibinfo {volume} {104}},\ \bibinfo {pages} {L201106} (\bibinfo {year} {2021})}\BibitemShut {NoStop}%
\bibitem [{\citenamefont {Sinha}\ \emph {et~al.}(2021)\citenamefont {Sinha}, \citenamefont {Chanda},\ and\ \citenamefont {Dziarmaga}}]{Sinha2021nonadiabiatic}%
  \BibitemOpen
  \bibfield  {author} {\bibinfo {author} {\bibfnamefont {A.}~\bibnamefont {Sinha}}, \bibinfo {author} {\bibfnamefont {T.}~\bibnamefont {Chanda}},\ and\ \bibinfo {author} {\bibfnamefont {J.}~\bibnamefont {Dziarmaga}},\ }\bibfield  {title} {\bibinfo {title} {Nonadiabatic dynamics across a first-order quantum phase transition: Quantized bubble nucleation},\ }\href {https://doi.org/10.1103/PhysRevB.103.L220302} {\bibfield  {journal} {\bibinfo  {journal} {Phys. Rev. B}\ }\textbf {\bibinfo {volume} {103}},\ \bibinfo {pages} {L220302} (\bibinfo {year} {2021})}\BibitemShut {NoStop}%
\bibitem [{\citenamefont {{D-Wave Systems}}(2020)}]{dwave2020TechnicalDescription}%
  \BibitemOpen
  \bibfield  {author} {\bibinfo {author} {\bibnamefont {{D-Wave Systems}}},\ }\href {https://docs.dwavesys.com/docs/latest/doc_qpu.html} {\emph {\bibinfo {title} {{Technical Description of the D-Wave Quantum Processing Unit}}}},\ \bibinfo {type} {Tech. Rep.}\ (\bibinfo  {institution} {D-Wave Systems Inc., Burnaby, BC, Canada},\ \bibinfo {year} {2020})\ \bibinfo {note} {{D-Wave User Manual 09-1109A-V}}\BibitemShut {NoStop}%
\bibitem [{\citenamefont {Amin}\ \emph {et~al.}(2008)\citenamefont {Amin}, \citenamefont {Love},\ and\ \citenamefont {Truncik}}]{amin2008thermally}%
  \BibitemOpen
  \bibfield  {author} {\bibinfo {author} {\bibfnamefont {M.~H.~S.}\ \bibnamefont {Amin}}, \bibinfo {author} {\bibfnamefont {P.~J.}\ \bibnamefont {Love}},\ and\ \bibinfo {author} {\bibfnamefont {C.~J.~S.}\ \bibnamefont {Truncik}},\ }\bibfield  {title} {\bibinfo {title} {Thermally assisted adiabatic quantum computation},\ }\href {https://doi.org/10.1103/PhysRevLett.100.060503} {\bibfield  {journal} {\bibinfo  {journal} {Phys. Rev. Lett.}\ }\textbf {\bibinfo {volume} {100}},\ \bibinfo {pages} {060503} (\bibinfo {year} {2008})}\BibitemShut {NoStop}%
\bibitem [{\citenamefont {Bravyi}\ \emph {et~al.}(2011)\citenamefont {Bravyi}, \citenamefont {DiVincenzo},\ and\ \citenamefont {Loss}}]{Bravyi2011}%
  \BibitemOpen
  \bibfield  {author} {\bibinfo {author} {\bibfnamefont {S.}~\bibnamefont {Bravyi}}, \bibinfo {author} {\bibfnamefont {D.~P.}\ \bibnamefont {DiVincenzo}},\ and\ \bibinfo {author} {\bibfnamefont {D.}~\bibnamefont {Loss}},\ }\bibfield  {title} {\bibinfo {title} {Schrieffer-wolff transformation for quantum many-body systems},\ }\href {https://doi.org/https://doi.org/10.1016/j.aop.2011.06.004} {\bibfield  {journal} {\bibinfo  {journal} {Annals of Physics}\ }\textbf {\bibinfo {volume} {326}},\ \bibinfo {pages} {2793 } (\bibinfo {year} {2011})}\BibitemShut {NoStop}%
\bibitem [{\citenamefont {Bernien}\ \emph {et~al.}(2017)\citenamefont {Bernien}, \citenamefont {Schwartz}, \citenamefont {Keesling}, \citenamefont {Levine}, \citenamefont {Omran}, \citenamefont {Pichler}, \citenamefont {Choi}, \citenamefont {Zibrov}, \citenamefont {Endres}, \citenamefont {Greiner} \emph {et~al.}}]{Bernien2017Rydberg}%
  \BibitemOpen
  \bibfield  {author} {\bibinfo {author} {\bibfnamefont {H.}~\bibnamefont {Bernien}}, \bibinfo {author} {\bibfnamefont {S.}~\bibnamefont {Schwartz}}, \bibinfo {author} {\bibfnamefont {A.}~\bibnamefont {Keesling}}, \bibinfo {author} {\bibfnamefont {H.}~\bibnamefont {Levine}}, \bibinfo {author} {\bibfnamefont {A.}~\bibnamefont {Omran}}, \bibinfo {author} {\bibfnamefont {H.}~\bibnamefont {Pichler}}, \bibinfo {author} {\bibfnamefont {S.}~\bibnamefont {Choi}}, \bibinfo {author} {\bibfnamefont {A.~S.}\ \bibnamefont {Zibrov}}, \bibinfo {author} {\bibfnamefont {M.}~\bibnamefont {Endres}}, \bibinfo {author} {\bibfnamefont {M.}~\bibnamefont {Greiner}}, \emph {et~al.},\ }\bibfield  {title} {\bibinfo {title} {Probing many-body dynamics on a 51-atom quantum simulator},\ }\href {https://doi.org/https://doi.org/10.1038/nature24622} {\bibfield  {journal} {\bibinfo  {journal} {Nature}\ }\textbf {\bibinfo {volume} {551}},\ \bibinfo {pages} {579} (\bibinfo {year} {2017})}\BibitemShut {NoStop}%
\bibitem [{\citenamefont {Polkovnikov}\ \emph {et~al.}(2011)\citenamefont {Polkovnikov}, \citenamefont {Sengupta}, \citenamefont {Silva},\ and\ \citenamefont {Vengalattore}}]{polkovnikov2011colloquium}%
  \BibitemOpen
  \bibfield  {author} {\bibinfo {author} {\bibfnamefont {A.}~\bibnamefont {Polkovnikov}}, \bibinfo {author} {\bibfnamefont {K.}~\bibnamefont {Sengupta}}, \bibinfo {author} {\bibfnamefont {A.}~\bibnamefont {Silva}},\ and\ \bibinfo {author} {\bibfnamefont {M.}~\bibnamefont {Vengalattore}},\ }\bibfield  {title} {\bibinfo {title} {Colloquium: Nonequilibrium dynamics of closed interacting quantum systems},\ }\href {https://doi.org/10.1103/RevModPhys.83.863} {\bibfield  {journal} {\bibinfo  {journal} {Rev. Mod. Phys.}\ }\textbf {\bibinfo {volume} {83}},\ \bibinfo {pages} {863} (\bibinfo {year} {2011})}\BibitemShut {NoStop}%
\bibitem [{\citenamefont {Kibble}(1976)}]{kibble1976topology}%
  \BibitemOpen
  \bibfield  {author} {\bibinfo {author} {\bibfnamefont {T.~W.}\ \bibnamefont {Kibble}},\ }\bibfield  {title} {\bibinfo {title} {Topology of cosmic domains and strings},\ }\href {https://doi.org/10.1088/0305-4470/9/8/029} {\bibfield  {journal} {\bibinfo  {journal} {Journal of Physics A: Mathematical and General}\ }\textbf {\bibinfo {volume} {9}},\ \bibinfo {pages} {1387} (\bibinfo {year} {1976})}\BibitemShut {NoStop}%
\bibitem [{\citenamefont {Dziarmaga}(2005)}]{dziarmaga2005dynamics}%
  \BibitemOpen
  \bibfield  {author} {\bibinfo {author} {\bibfnamefont {J.}~\bibnamefont {Dziarmaga}},\ }\bibfield  {title} {\bibinfo {title} {Dynamics of a quantum phase transition: Exact solution of the quantum {Ising} model},\ }\href {https://doi.org/10.1103/PhysRevLett.95.245701} {\bibfield  {journal} {\bibinfo  {journal} {Phys. Rev. Lett.}\ }\textbf {\bibinfo {volume} {95}},\ \bibinfo {pages} {245701} (\bibinfo {year} {2005})}\BibitemShut {NoStop}%
\bibitem [{\citenamefont {Langer}(1969)}]{langer1969statistical}%
  \BibitemOpen
  \bibfield  {author} {\bibinfo {author} {\bibfnamefont {J.~S.}\ \bibnamefont {Langer}},\ }\bibfield  {title} {\bibinfo {title} {Statistical theory of the decay of metastable states},\ }\href {https://doi.org/10.1016/0003-4916(69)90153-5} {\bibfield  {journal} {\bibinfo  {journal} {Annals of Physics}\ }\textbf {\bibinfo {volume} {54}},\ \bibinfo {pages} {258} (\bibinfo {year} {1969})}\BibitemShut {NoStop}%
\bibitem [{\citenamefont {Affleck}(1981)}]{affleck1981quantum}%
  \BibitemOpen
  \bibfield  {author} {\bibinfo {author} {\bibfnamefont {I.}~\bibnamefont {Affleck}},\ }\bibfield  {title} {\bibinfo {title} {Quantum-statistical metastability},\ }\href {https://doi.org/10.1103/PhysRevLett.46.388} {\bibfield  {journal} {\bibinfo  {journal} {Phys. Rev. Lett.}\ }\textbf {\bibinfo {volume} {46}},\ \bibinfo {pages} {388} (\bibinfo {year} {1981})}\BibitemShut {NoStop}%
\bibitem [{\citenamefont {Caldeira}\ and\ \citenamefont {Leggett}(1981)}]{caldeira1981influence}%
  \BibitemOpen
  \bibfield  {author} {\bibinfo {author} {\bibfnamefont {A.~O.}\ \bibnamefont {Caldeira}}\ and\ \bibinfo {author} {\bibfnamefont {A.~J.}\ \bibnamefont {Leggett}},\ }\bibfield  {title} {\bibinfo {title} {Influence of dissipation on quantum tunneling in macroscopic systems},\ }\href {https://doi.org/10.1103/PhysRevLett.46.211} {\bibfield  {journal} {\bibinfo  {journal} {Phys. Rev. Lett.}\ }\textbf {\bibinfo {volume} {46}},\ \bibinfo {pages} {211} (\bibinfo {year} {1981})}\BibitemShut {NoStop}%
\bibitem [{\citenamefont {Leggett}(1984)}]{leggett1984quantum}%
  \BibitemOpen
  \bibfield  {author} {\bibinfo {author} {\bibfnamefont {A.~J.}\ \bibnamefont {Leggett}},\ }\bibfield  {title} {\bibinfo {title} {Quantum tunneling in the presence of an arbitrary linear dissipation mechanism},\ }\href {https://doi.org/10.1103/PhysRevB.30.1208} {\bibfield  {journal} {\bibinfo  {journal} {Phys. Rev. B}\ }\textbf {\bibinfo {volume} {30}},\ \bibinfo {pages} {1208} (\bibinfo {year} {1984})}\BibitemShut {NoStop}%
\bibitem [{\citenamefont {Leggett}\ \emph {et~al.}(1987)\citenamefont {Leggett}, \citenamefont {Chakravarty}, \citenamefont {Dorsey}, \citenamefont {Fisher}, \citenamefont {Garg},\ and\ \citenamefont {Zwerger}}]{leggett1987dynamics}%
  \BibitemOpen
  \bibfield  {author} {\bibinfo {author} {\bibfnamefont {A.~J.}\ \bibnamefont {Leggett}}, \bibinfo {author} {\bibfnamefont {S.}~\bibnamefont {Chakravarty}}, \bibinfo {author} {\bibfnamefont {A.~T.}\ \bibnamefont {Dorsey}}, \bibinfo {author} {\bibfnamefont {M.~P.~A.}\ \bibnamefont {Fisher}}, \bibinfo {author} {\bibfnamefont {A.}~\bibnamefont {Garg}},\ and\ \bibinfo {author} {\bibfnamefont {W.}~\bibnamefont {Zwerger}},\ }\bibfield  {title} {\bibinfo {title} {Dynamics of the dissipative two-state system},\ }\href {https://doi.org/10.1103/RevModPhys.59.1} {\bibfield  {journal} {\bibinfo  {journal} {Rev. Mod. Phys.}\ }\textbf {\bibinfo {volume} {59}},\ \bibinfo {pages} {1} (\bibinfo {year} {1987})}\BibitemShut {NoStop}%
\bibitem [{\citenamefont {H\"anggi}\ \emph {et~al.}(1990)\citenamefont {H\"anggi}, \citenamefont {Talkner},\ and\ \citenamefont {Borkovec}}]{hanggi1990reaction}%
  \BibitemOpen
  \bibfield  {author} {\bibinfo {author} {\bibfnamefont {P.}~\bibnamefont {H\"anggi}}, \bibinfo {author} {\bibfnamefont {P.}~\bibnamefont {Talkner}},\ and\ \bibinfo {author} {\bibfnamefont {M.}~\bibnamefont {Borkovec}},\ }\bibfield  {title} {\bibinfo {title} {Reaction-rate theory: fifty years after {Kramers}},\ }\href {https://doi.org/10.1103/RevModPhys.62.251} {\bibfield  {journal} {\bibinfo  {journal} {Rev. Mod. Phys.}\ }\textbf {\bibinfo {volume} {62}},\ \bibinfo {pages} {251} (\bibinfo {year} {1990})}\BibitemShut {NoStop}%
\bibitem [{\citenamefont {Birnkammer}\ \emph {et~al.}(2022)\citenamefont {Birnkammer}, \citenamefont {Bastianello},\ and\ \citenamefont {Knap}}]{birnkammer2022prethermalization}%
  \BibitemOpen
  \bibfield  {author} {\bibinfo {author} {\bibfnamefont {S.}~\bibnamefont {Birnkammer}}, \bibinfo {author} {\bibfnamefont {A.}~\bibnamefont {Bastianello}},\ and\ \bibinfo {author} {\bibfnamefont {M.}~\bibnamefont {Knap}},\ }\bibfield  {title} {\bibinfo {title} {Prethermalization in one-dimensional quantum many-body systems with confinement},\ }\href {https://doi.org/10.1038/s41467-022-35301-6} {\bibfield  {journal} {\bibinfo  {journal} {Nat. Commun.}\ }\textbf {\bibinfo {volume} {13}},\ \bibinfo {pages} {7663} (\bibinfo {year} {2022})}\BibitemShut {NoStop}%
\bibitem [{\citenamefont {Fendley}\ \emph {et~al.}(2004)\citenamefont {Fendley}, \citenamefont {Sengupta},\ and\ \citenamefont {Sachdev}}]{FendleySachdev}%
  \BibitemOpen
  \bibfield  {author} {\bibinfo {author} {\bibfnamefont {P.}~\bibnamefont {Fendley}}, \bibinfo {author} {\bibfnamefont {K.}~\bibnamefont {Sengupta}},\ and\ \bibinfo {author} {\bibfnamefont {S.}~\bibnamefont {Sachdev}},\ }\bibfield  {title} {\bibinfo {title} {Competing density-wave orders in a one-dimensional hard-boson model},\ }\href {https://doi.org/10.1103/PhysRevB.69.075106} {\bibfield  {journal} {\bibinfo  {journal} {Phys. Rev. B}\ }\textbf {\bibinfo {volume} {69}},\ \bibinfo {pages} {075106} (\bibinfo {year} {2004})}\BibitemShut {NoStop}%
\bibitem [{\citenamefont {Lesanovsky}\ and\ \citenamefont {Katsura}(2012)}]{Lesanovsky2012}%
  \BibitemOpen
  \bibfield  {author} {\bibinfo {author} {\bibfnamefont {I.}~\bibnamefont {Lesanovsky}}\ and\ \bibinfo {author} {\bibfnamefont {H.}~\bibnamefont {Katsura}},\ }\bibfield  {title} {\bibinfo {title} {Interacting {Fibonacci} anyons in a {Rydberg} gas},\ }\href {https://doi.org/10.1103/PhysRevA.86.041601} {\bibfield  {journal} {\bibinfo  {journal} {Phys. Rev. A}\ }\textbf {\bibinfo {volume} {86}},\ \bibinfo {pages} {041601(R)} (\bibinfo {year} {2012})}\BibitemShut {NoStop}%
\bibitem [{\citenamefont {Turner}\ \emph {et~al.}(2018)\citenamefont {Turner}, \citenamefont {Michailidis}, \citenamefont {Abanin}, \citenamefont {Serbyn},\ and\ \citenamefont {Papi{\'c}}}]{TurnerNature}%
  \BibitemOpen
  \bibfield  {author} {\bibinfo {author} {\bibfnamefont {C.~J.}\ \bibnamefont {Turner}}, \bibinfo {author} {\bibfnamefont {A.~A.}\ \bibnamefont {Michailidis}}, \bibinfo {author} {\bibfnamefont {D.~A.}\ \bibnamefont {Abanin}}, \bibinfo {author} {\bibfnamefont {M.}~\bibnamefont {Serbyn}},\ and\ \bibinfo {author} {\bibfnamefont {Z.}~\bibnamefont {Papi{\'c}}},\ }\bibfield  {title} {\bibinfo {title} {Weak ergodicity breaking from quantum many-body scars},\ }\href {https://doi.org/https://doi.org/10.1038/s41567-018-0137-5} {\bibfield  {journal} {\bibinfo  {journal} {Nat. Phys.}\ }\textbf {\bibinfo {volume} {14}},\ \bibinfo {pages} {745} (\bibinfo {year} {2018})}\BibitemShut {NoStop}%
\bibitem [{\citenamefont {Su}\ \emph {et~al.}(2023)\citenamefont {Su}, \citenamefont {Sun}, \citenamefont {Hudomal}, \citenamefont {Desaules}, \citenamefont {Zhou}, \citenamefont {Yang}, \citenamefont {Halimeh}, \citenamefont {Yuan}, \citenamefont {Papi\ifmmode~\acute{c}\else \'{c}\fi{}},\ and\ \citenamefont {Pan}}]{Su2023TBH}%
  \BibitemOpen
  \bibfield  {author} {\bibinfo {author} {\bibfnamefont {G.-X.}\ \bibnamefont {Su}}, \bibinfo {author} {\bibfnamefont {H.}~\bibnamefont {Sun}}, \bibinfo {author} {\bibfnamefont {A.}~\bibnamefont {Hudomal}}, \bibinfo {author} {\bibfnamefont {J.-Y.}\ \bibnamefont {Desaules}}, \bibinfo {author} {\bibfnamefont {Z.-Y.}\ \bibnamefont {Zhou}}, \bibinfo {author} {\bibfnamefont {B.}~\bibnamefont {Yang}}, \bibinfo {author} {\bibfnamefont {J.~C.}\ \bibnamefont {Halimeh}}, \bibinfo {author} {\bibfnamefont {Z.-S.}\ \bibnamefont {Yuan}}, \bibinfo {author} {\bibfnamefont {Z.}~\bibnamefont {Papi\ifmmode~\acute{c}\else \'{c}\fi{}}},\ and\ \bibinfo {author} {\bibfnamefont {J.-W.}\ \bibnamefont {Pan}},\ }\bibfield  {title} {\bibinfo {title} {Observation of many-body scarring in a {Bose}-{Hubbard} quantum simulator},\ }\href {https://doi.org/10.1103/PhysRevResearch.5.023010} {\bibfield  {journal} {\bibinfo  {journal} {Phys. Rev. Res.}\ }\textbf {\bibinfo {volume} {5}},\ \bibinfo {pages} {023010} (\bibinfo {year}
  {2023})}\BibitemShut {NoStop}%
\bibitem [{\citenamefont {Balducci}\ \emph {et~al.}(2022)\citenamefont {Balducci}, \citenamefont {Gambassi}, \citenamefont {Lerose}, \citenamefont {Scardicchio},\ and\ \citenamefont {Vanoni}}]{balducci2022localization}%
  \BibitemOpen
  \bibfield  {author} {\bibinfo {author} {\bibfnamefont {F.}~\bibnamefont {Balducci}}, \bibinfo {author} {\bibfnamefont {A.}~\bibnamefont {Gambassi}}, \bibinfo {author} {\bibfnamefont {A.}~\bibnamefont {Lerose}}, \bibinfo {author} {\bibfnamefont {A.}~\bibnamefont {Scardicchio}},\ and\ \bibinfo {author} {\bibfnamefont {C.}~\bibnamefont {Vanoni}},\ }\bibfield  {title} {\bibinfo {title} {Localization and melting of interfaces in the two-dimensional quantum ising model},\ }\href {https://doi.org/10.1103/PhysRevLett.129.120601} {\bibfield  {journal} {\bibinfo  {journal} {Phys. Rev. Lett.}\ }\textbf {\bibinfo {volume} {129}},\ \bibinfo {pages} {120601} (\bibinfo {year} {2022})}\BibitemShut {NoStop}%
\bibitem [{\citenamefont {Hart}\ and\ \citenamefont {Nandkishore}(2022)}]{Hart2022}%
  \BibitemOpen
  \bibfield  {author} {\bibinfo {author} {\bibfnamefont {O.}~\bibnamefont {Hart}}\ and\ \bibinfo {author} {\bibfnamefont {R.}~\bibnamefont {Nandkishore}},\ }\bibfield  {title} {\bibinfo {title} {Hilbert space shattering and dynamical freezing in the quantum ising model},\ }\href {https://doi.org/10.1103/PhysRevB.106.214426} {\bibfield  {journal} {\bibinfo  {journal} {Phys. Rev. B}\ }\textbf {\bibinfo {volume} {106}},\ \bibinfo {pages} {214426} (\bibinfo {year} {2022})}\BibitemShut {NoStop}%
\bibitem [{\citenamefont {Karp}(1972)}]{Karp1972KarpsNPCompleteProblems}%
  \BibitemOpen
  \bibfield  {author} {\bibinfo {author} {\bibfnamefont {R.~M.}\ \bibnamefont {Karp}},\ }\bibinfo {title} {Reducibility among combinatorial problems},\ in\ \href {https://doi.org/10.1007/978-1-4684-2001-2_9} {\emph {\bibinfo {booktitle} {Complexity of Computer Computations}}},\ \bibinfo {series and number} {The IBM Research Symposia Series},\ \bibinfo {editor} {edited by\ \bibinfo {editor} {\bibfnamefont {R.~E.}\ \bibnamefont {Miller}}, \bibinfo {editor} {\bibfnamefont {J.~W.}\ \bibnamefont {Thatcher}},\ and\ \bibinfo {editor} {\bibfnamefont {J.~D.}\ \bibnamefont {Bohlinger}}}\ (\bibinfo  {publisher} {Springer US},\ \bibinfo {address} {Boston, MA},\ \bibinfo {year} {1972})\ pp.\ \bibinfo {pages} {85--103}\BibitemShut {NoStop}%
\bibitem [{\citenamefont {Breuer}\ and\ \citenamefont {Petruccione}(2007)}]{breuer2002theory}%
  \BibitemOpen
  \bibfield  {author} {\bibinfo {author} {\bibfnamefont {H.-P.}\ \bibnamefont {Breuer}}\ and\ \bibinfo {author} {\bibfnamefont {F.}~\bibnamefont {Petruccione}},\ }\href {https://doi.org/10.1093/acprof:oso/9780199213900.001.0001} {\emph {\bibinfo {title} {The Theory of Open Quantum Systems}}}\ (\bibinfo  {publisher} {Oxford University Press},\ \bibinfo {year} {2007})\BibitemShut {NoStop}%
\bibitem [{\citenamefont {Schollw{\"o}ck}(2011)}]{Schollwock}%
  \BibitemOpen
  \bibfield  {author} {\bibinfo {author} {\bibfnamefont {U.}~\bibnamefont {Schollw{\"o}ck}},\ }\bibfield  {title} {\bibinfo {title} {The density-matrix renormalization group in the age of matrix product states},\ }\href {https://doi.org/https://doi.org/10.1016/j.aop.2010.09.012} {\bibfield  {journal} {\bibinfo  {journal} {Annals of Physics}\ }\textbf {\bibinfo {volume} {326}},\ \bibinfo {pages} {96 } (\bibinfo {year} {2011})}\BibitemShut {NoStop}%
\bibitem [{\citenamefont {Vidal}(2003)}]{Vidal2003}%
  \BibitemOpen
  \bibfield  {author} {\bibinfo {author} {\bibfnamefont {G.}~\bibnamefont {Vidal}},\ }\bibfield  {title} {\bibinfo {title} {Efficient classical simulation of slightly entangled quantum computations},\ }\href {https://doi.org/10.1103/PhysRevLett.91.147902} {\bibfield  {journal} {\bibinfo  {journal} {Phys. Rev. Lett.}\ }\textbf {\bibinfo {volume} {91}},\ \bibinfo {pages} {147902} (\bibinfo {year} {2003})}\BibitemShut {NoStop}%
\bibitem [{\citenamefont {Paeckel}\ \emph {et~al.}(2019)\citenamefont {Paeckel}, \citenamefont {Köhler}, \citenamefont {Swoboda}, \citenamefont {Manmana}, \citenamefont {Schollwöck},\ and\ \citenamefont {Hubig}}]{Paeckel_2019}%
  \BibitemOpen
  \bibfield  {author} {\bibinfo {author} {\bibfnamefont {S.}~\bibnamefont {Paeckel}}, \bibinfo {author} {\bibfnamefont {T.}~\bibnamefont {Köhler}}, \bibinfo {author} {\bibfnamefont {A.}~\bibnamefont {Swoboda}}, \bibinfo {author} {\bibfnamefont {S.~R.}\ \bibnamefont {Manmana}}, \bibinfo {author} {\bibfnamefont {U.}~\bibnamefont {Schollwöck}},\ and\ \bibinfo {author} {\bibfnamefont {C.}~\bibnamefont {Hubig}},\ }\bibfield  {title} {\bibinfo {title} {Time-evolution methods for matrix-product states},\ }\href {https://doi.org/10.1016/j.aop.2019.167998} {\bibfield  {journal} {\bibinfo  {journal} {Annals of Physics}\ }\textbf {\bibinfo {volume} {411}},\ \bibinfo {pages} {167998} (\bibinfo {year} {2019})}\BibitemShut {NoStop}%
\end{thebibliography}%

\end{document}